\documentclass[10pt, journal]{IEEEtran}
\IEEEoverridecommandlockouts
\usepackage{cite}
\usepackage{amsmath,amssymb,amsfonts}
\usepackage{comment}
\usepackage{algorithmic}
\usepackage{graphicx}
\usepackage{textcomp}
\usepackage{xcolor}
\usepackage[pscoord]{eso-pic}
\usepackage{balance}
\usepackage{fancyvrb}
\usepackage{acronym}
\acrodef{IP}[IP]{intellectual property block}
\acrodef{SoC}[SoC]{System-on-Chip}
\acrodef{IC}[IC]{integrated circuit}
\acrodef{eFPGA}[eFPGA]{embedded field programmable gate array}
\acrodef{FPGA}[FPGA]{field programmable gate array}
\acrodef{RTL}[RTL]{register transfer level}
\acrodef{CLB}[CLB]{configurable logic block}
\acrodef{LUT}[LUT]{look-up table}
\acrodef{HLS}[HLS]{high-level synthesis}
\acrodef{EDA}[EDA]{electronic design automation}
\acrodef{FF}[FF]{flip-flop}
\acrodef{DIP}[DIP]{distinguishing input pattern}
\acrodef{PPA}[PPA]{power, performance, and area}
\acrodef{BLE}[BLE]{basic logic element}
\acrodef{CB}[CB]{connection block}
\acrodef{SB}[SB]{switch block}

\usepackage{diagbox}

\usepackage{tabularx}
\newcolumntype{L}[1]{>{\raggedright\let\newline\\\arraybackslash\hspace{0pt}}m{#1}}
\newcolumntype{C}[1]{>{\centering\let\newline\\\arraybackslash\hspace{0pt}}m{#1}}
\newcolumntype{R}[1]{>{\raggedleft\let\newline\\\arraybackslash\hspace{0pt}}m{#1}}

\def\BibTeX{{\rm B\kern-.05em{\sc i\kern-.025em b}\kern-.08em
    T\kern-.1667em\lower.7ex\hbox{E}\kern-.125emX}}

\def\BibTeX{{\rm B\kern-.05em{\sc i\kern-.025em b}\kern-.08em
    T\kern-.1667em\lower.7ex\hbox{E}\kern-.125emX}}
\def\BibTeX{{\rm B\kern-.05em{\sc i\kern-.025em b}\kern-.08em
    T\kern-.1667em\lower.7ex\hbox{E}\kern-.125emX}}

\usepackage{pifont}
\usepackage{framed}
\usepackage{soul}
\usepackage{booktabs}  
\usepackage{csquotes}
\usepackage{multirow}
\usepackage{array}
\newcolumntype{L}[1]{>{\raggedright\let\newline\\\arraybackslash\hspace{0pt}}m{#1}}
\newcolumntype{C}[1]{>{\centering\let\newline\\\arraybackslash\hspace{0pt}}m{#1}}
\newcolumntype{R}[1]{>{\raggedleft\let\newline\\\arraybackslash\hspace{0pt}}m{#1}}
\usepackage{url}
\MakeOuterQuote{"}
\usepackage[caption=false,font=footnotesize,labelformat=simple]{subfig} 

\newcommand{\todoblockblue}[1]{\noindent\definecolor{shadecolor}{RGB}{214, 254, 255}\colorbox{shadecolor}{\parbox{\columnwidth}{{#1}}}}

\usepackage[bookmarks=false]{hyperref}

\newcommand{\cmark}{\ding{51}}%
\newcommand{\xmark}{\ding{55}}%

\begin{document}

\bstctlcite{IEEEexample:BSTcontrol}

\title{%
Not All Fabrics Are Created Equal: Exploring eFPGA Parameters For IP Redaction
}

\author{Jitendra Bhandari, 
Abdul Khader Thalakkattu Moosa, 
Benjamin Tan,~\IEEEmembership{Member,~IEEE},\\ 
Christian Pilato,~\IEEEmembership{Senior~Member,~IEEE},
Ganesh Gore, 
Xifan Tang,~\IEEEmembership{Member,~IEEE},
Scott Temple,\\ 
Pierre-Emmanuel Gaillardon,~\IEEEmembership{Senior~Member,~IEEE},
and Ramesh Karri~\IEEEmembership{Fellow,~IEEE}
\thanks{This work was supported in part by NYU CCS. Ganesh Gore, Xifan Tang, Pierre-Emmanuel Gaillardon are supported by AFRL and DARPA under agreement number FA8650-18-2-7855, and Scott Temple, Pierre-Emmanuel Gaillardon are supported by AFRL and DARPA under agreement number FA8650-18-2-7849.\protect\\ 
\indent J. Bhandari and A. Khader Thalakkattu Moosa contributed equally to this work.\protect\\
\indent J. Bhandari, A. Khader Thalakkattu Moosa, and R. Karri are with the Center for Cybersecurity, New York University, New York City, NY, 11201 USA. E-mail: \{jb7410, at4856, rkarri\}@nyu.edu \protect\\
\indent B. Tan is with the Department of Electrical and Software Engineering, University of Calgary, Calgary, AB, T2N 1N4, Canada. Email: benjamin.tan1@ucalgary.ca \protect\\
\indent C. Pilato is with the Dipartimento di Elettronica, Informazione e Bioingegneria, Politecnico di Milano, Milano, 20133, Italy. Email: christian.pilato@polimi.it \protect\\
\indent G. Gore, X. Tang, S. Temple, P.-E. Gaillardon are with the Laboratory for NanoIntegrated Systems, Electrical and Computer Engineering Department, University of Utah, Salt Lake City, UT 84112 USA. E-mail: \{ganesh.gore, xifan.tang, scott.temple, pierre-emmanuel.gaillardon\}@utah.edu
}
}

\maketitle

\begin{abstract}
Semiconductor design houses rely on third-party foundries to manufacture their integrated circuits (IC). While this trend allows them to tackle fabrication costs, it introduces security concerns as external (and potentially malicious) parties can access critical parts of the designs and steal or modify the Intellectual Property (IP). 
Embedded FPGA (eFPGA) redaction is a promising technique to protect critical IPs of an ASIC by \textit{redacting} (i.e., removing) critical parts and mapping them onto a custom reconfigurable fabric. Only trusted parties will receive the correct bitstream to restore the redacted functionality.
While previous studies imply that using an eFPGA is a sufficient condition to provide security against IP threats like reverse-engineering, whether this truly holds for all eFPGA architectures is unclear, thus motivating the study in this paper. 
We examine the security of eFPGA fabrics generated by varying different FPGA design parameters. We characterize the power, performance, and area (PPA) characteristics and evaluate each fabric's resistance to SAT-based bitstream recovery. Our results encourage designers to work with custom eFPGA fabrics rather than off-the-shelf commercial FPGAs and reveals that only considering a redaction fabric's bitstream size is inadequate for gauging security. 

\end{abstract}


\section{Introduction}
\begin{figure}[t]
\centering
\subfloat[\label{fig:ic_supply}]{\includegraphics[width=1\columnwidth]{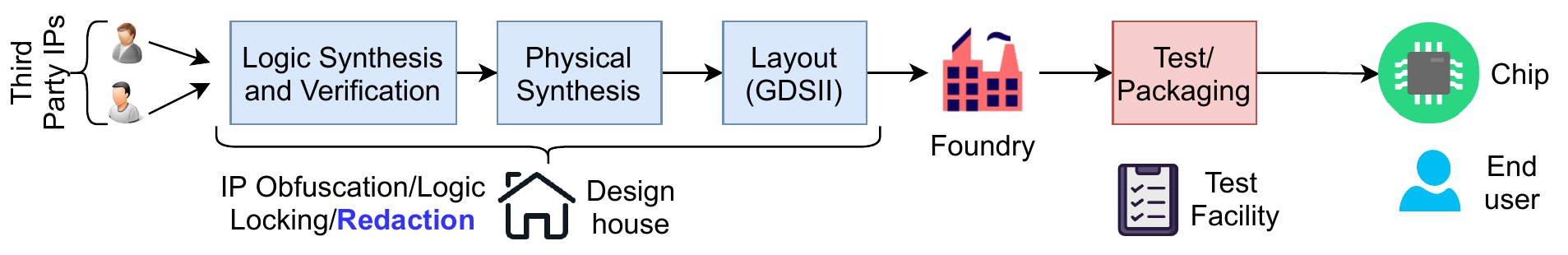}} 
\\
\subfloat[\label{fig:redact_over}]{\includegraphics[width=0.5\columnwidth]{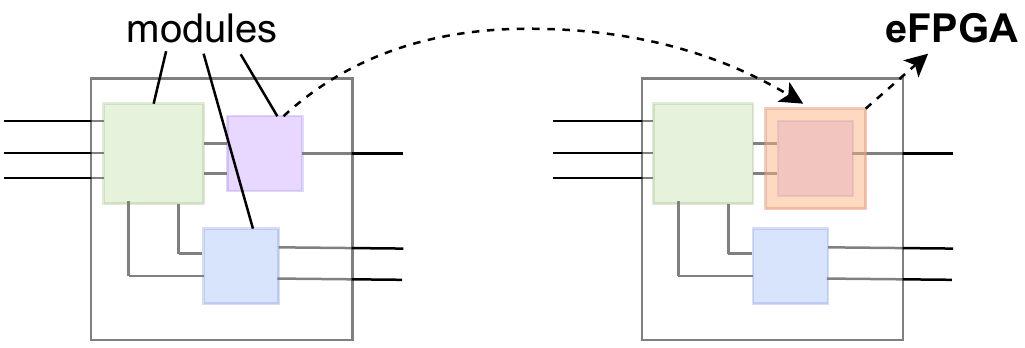}}

\caption{(a) \ac{IC} design and fabrication flow and possible way to secure in the design house stage. (b) eFPGA redaction takes a module of an IP and implements it as a reprogrammable fabric replacing the redacted function.}
\label{fig:supply_chain}
\end{figure}

As technology advances, \ac{IC} complexity has grown significantly and led to increased outsourcing of the steps of the design flow to third-party entities in the supply-chain, as shown in \autoref{fig:ic_supply}. Outsourcing and globalization introduce many players in the supply-chain and this presents challenges of \ac{IP} theft, reverse-engineering, and malicious manipulation~\cite{rostami_primer_2014}. 
Consider \ac{IC} layout design files which are sent to the foundry for fabrication, malicious (or compromised) employees can access these files and reverse-engineer the function to steal the \ac{IP} of critical design portions or to insert hardware Trojans. Malicious end-users can obtain working ICs to analyze the I/O relationships and reverse-engineer the correct function (in collusion with a malicious foundry) to make unauthorized clones.

In response,
researchers have proposed a myriad of solutions that aim to protect the confidentiality of the hardware \ac{IP}, including design obfuscation and logic locking (e.g., ~\cite{pilato_assure_2021,LLC,shamsi_ip_2019, mohan_hardware_2021, chen_decoy_2020, hu_functional_2019,kamali_interlock_2020,kolhe_security_2019, liu_embedded_2014,limaye_thwarting_2020}). 
All these techniques obscure a design's function by adding modules whose correct functionality depends on an external \textit{key}~\cite{limaye_thwarting_2020} or by withholding information such as algorithm constants that only a legitimate user can later restore~\cite{pilato_assure_2021}. 
Incorrect keys corrupt the \ac{IC}'s functionality, rendering the design useless to the malicious party. 
To date, attacks have overcome the protections. The most notable class of attacks is based on Boolean satisfiability (SAT)~\cite{subramanyan_evaluating_2015,LLC,shamsi_icysat_2019}. These SAT-based attacks assume an adversary with access to an unlocked implementation (the \textit{Oracle}).  
Recently, redacting parts of an \ac{IP} by using \acp{eFPGA}, as depicted in \autoref{fig:redact_over}, has emerged as a promising, SAT-attack resilient defense~\cite{mohan_hardware_2021,hu_functional_2019}. 
The intuition is that even small \acp{eFPGA} fabrics are insurmountable for SAT solvers because of their complexity and their size when converted into a representation for SAT solving~\cite{mohan_hardware_2021}. 
Prior work has begun to characterize the feasibility of this defense by studying the overhead associated with this technique assuming a fixed eFPGA architecture~\cite{our_iccad}.

\textbf{However, are all \ac{eFPGA}s the same from a security perspective?} 
To the best of our knowledge, the literature does not yet offer insights on how different \ac{eFPGA} parameters, such as look-up table size, affect the security offered by eFPGA-based redaction. 
In logic locking approaches, security is distilled into a single parameter: the \textbf{key size}~\cite{LLC} -- a designer can choose a key size, as a measure of security, and incur follow-on impacts on power, performance, and hardware area (PPA) metrics. In eFPGAs, the counterpart is the configuration \textbf{bitstream size}, which is determined by the conflation of \textit{multitudinous} design choices, from logic element configuration through to routing channel width (see \autoref{sec:background}) -- in other words, the \ac{eFPGA} design space is vast~\cite{FPGA_arch}. 
From a practical standpoint, it is crucial for designers to understand the relationships between security and other design factors.  Thus, we address this gap in literature by performing an \textbf{empirical study of eFPGA architecture configurations and resistance to bitstream recovery through SAT-attack as the security metric}. For insights into eFPGA-based IP redaction, we adapt an open-source \acs{FPGA} design flow~\cite{tang_openfpga_2020} to produce different \acp{eFPGA} fabrics, with different configurations, and explore how eFPGA parameters affect security. 
Our contributions are threefold: 




\begin{enumerate}
    \item An analysis of eFPGA architectures that can be used for redaction. We analyze PPA and security effects and explore how the parameter choices of an eFPGA fabric ``contribute'' to the security provided by it. 
    \item A formulation of SAT-based attack for bitstream recovery of eFPGAs used for redaction and an experimental evaluation of eFPGA-based defense.
    \item Insights into the practical considerations for adopting eFPGA-based redaction and a perspective on the future outlook of this IP protection technique.
\end{enumerate}
%
In \autoref{sec:2_prelims}, we present the hardware IP protection problem  alongside prior work to tackle this issue. This is followed by an introduction to eFPGA and their architectural parameters in \autoref{sec:background}, and architecture settings that we explore. 
\autoref{sec:security-I} details our initial attempts at eFPGA bitstream recovery, including threat model and assumptions.  \autoref{sec:security-II}  revises our approach for eFPGA bitstream recovery, with more insights into the security of eFPGA fabrics. 
We discuss the insights from our study in \autoref{sec:discussion} and then conclude in \autoref{sec:conclusions}.

\section{Related Work and Motivation\label{sec:2_prelims}}

\subsection{Key-based Hardware IP Protection}

Logic locking is a popular technique for hardware \ac{IP} protection~\cite{LLC,shamsi_ip_2019}. Designers insert additional gates (controlled by an input key) to thwart reverse engineering of the real functionality. The key is known to the design house but unknown to the foundry. The correct key is installed into the chip after fabrication, assuming it is the only one that restores the correct functionality. So, \textbf{the key is the one and only secret to be protected (by the designers) or retrieved (by the attackers).} 
Attackers may also have access to a working chip (called \textit{Oracle}) to attempt key recovery by analyzing I/O relationships with SAT-based formulations~\cite{subramanyan_evaluating_2015,Azar_Kamali_Homayoun_Sasan_2018,shamsi_icysat_2019,be_sat,cyc_sat,shamsi_kc2_2019}. Otherwise, they can analyze the existence of structural artifacts~\cite{li_piercing_2019,han_does_2021} to guess the correct key bits.
On the other hand, designers need to design their locking techniques such that they 1) protect the essential semantics of the circuit, 2) guarantee that the key is not easy to be retrieved, and 3) minimize the hardware overhead.

For protecting essential semantics, locking is applied at \ac{RTL}, even though these methods incur in significant area overhead~\cite{pilato_assure_2021}. Other methods aim at trading off different security metrics, like SAT resilience and corruptibility~\cite{Shakya_Xu_Tehranipoor_Forte_2019}, but these approaches have structural vulnerabilities, leading to key recovery~\cite{Sengupta_Limaye_Sinanoglu_2021}. In all cases, the security of such key-locked design is proportional to the number of key bits. However, the key cannot grow indefinitely because of technological constraints like the size of the tamper-proof memory where it is installed.

\begin{figure*}[t]
    \centering
    \includegraphics[width=0.9\textwidth]{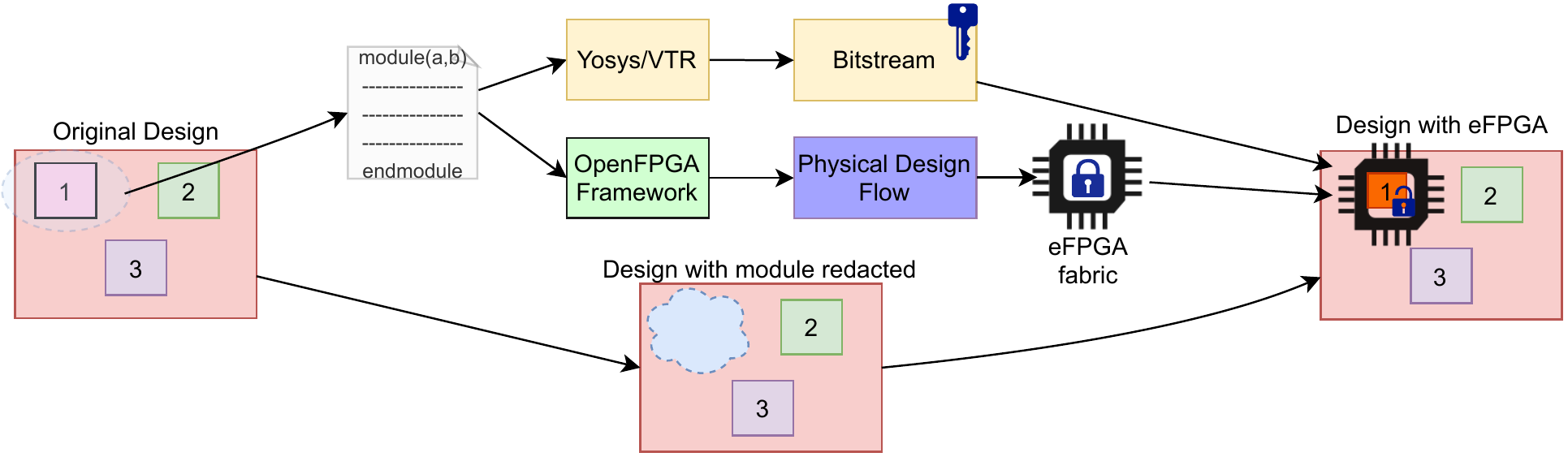}
    \caption{An \ac{eFPGA}-based redaction flow for \ac{RTL} \ac{IP}. The redacted portion (module) is picked by a designer. We adapt the OpenFPGA flow to produce the required eFPGA fabric, which we treat as a macro and connect to the remaining portion of the design. }
    \label{fig:redact-flow}
\end{figure*}

\subsection{eFPGA-based Redaction}

IP redaction is an alternative method to logic obfuscation. In this case, designers select specific modules -- the ones they want to protect -- to conceal and replace them with \textit{soft \acp{eFPGA}} (i.e., reconfigurable fabrics described in RTL and designed using standard-cells with the rest of the chip). The key idea is that only a sub-design gives the design house a competitive market advantage. An \ac{eFPGA} is a soft IP module that includes \acp{CLB} containing \acp{LUT}, flip-flops, and routing logic that can be fabricated and programmed to implement the desired functionality. The specific configuration of such devices is called \textit{bitstream}. A bitstream must include the configuration of each configurable module of the \ac{eFPGA}. When used for redaction, the designer will insert the \ac{eFPGA} module to replace the ``sensitive'' parts of the design that are thus unknown to untrusted parties during fabrication. On the contrary, the attacker must recover the complete bitstream to implement the correct functionality in each \ac{eFPGA}, which is now the ``secret'' to be protected.
\ac{eFPGA}-based redaction is particularly attracting for thwarting several reverse-engineering attacks. On one hand, structural attacks are difficult to be applied because the \ac{eFPGA} is regular and generic, able to implement an arbitrary functionality. On the other hand, the size of the configuration bitstream grows exponentially with the complexity of the \ac{eFPGA} architecture, significantly enlarging the key space and so thwarting SAT-based attacks~\cite{hu_functional_2019,mohan_hardware_2021,our_iccad}.

\autoref{fig:redact-flow} shows the general flow for \ac{eFPGA}-based redaction. After selecting the portion of the design to be redact, it goes through the \textit{fabric generation step}, while the rest of the chip is designed and optimized as usual. The fabric netlist describes the \ac{eFPGA} architecture and is then recombined with the rest of the chip to go through the physical design flow. The redacted module is instead compiled into a bitstream to configure the  \ac{eFPGA} after manufacturing.
%
While the approach is promising, there are several open issues to be addressed. \textbf{Which module(s) should a designer redact? What is the impact of inserting \acp{eFPGA} into the ASIC design flow? How can the designer generate the proper \ac{eFPGA} architecture? Are all \ac{eFPGA} architectures equally secure?}

When deciding which module(s) to redact, the designer could know the ``sensitive'' parts of the design, manually driving the selection~\cite{mohan_hardware_2021}, or use methods based on \ac{HLS} to identify the logic that differentiates variants of the same design~\cite{chen_decoy_2020}. In all cases, the designer assumes a standard or even off-the-shelf implementation of the \ac{eFPGA}, incurring in significant overheads. 
Several open-source and complete CAD flows can be used to generate the \ac{eFPGA} architectures and the corresponding bitstream for the modules to be redacted. Yosys and VTR/VPR can be used to identify the fabric parameters, along with a Chisel-based generator~\cite{mohan_hardware_2021}. OpenFPGA is an open-source generator of highly-customizable FPGA architectures that can be combined with the associated logic synthesis flows to generate the proper configuration bit-stream~\cite{tang_openfpga_2019}. While these approaches allow the exploration of different fabric parameters, their security and design implications have never been adequately explored in the case of IP redaction.
The security of \ac{eFPGA}-based redaction comes, in principle, with the size of the configuration bitstream rendering traditional SAT-based attacks infeasible~\cite{mohan_hardware_2021}. 
However, \textbf{the impact of different \ac{eFPGA} architectures on SAT resilience is unclear}; exploring this impact is the topic of our work. 

We base our exploration on OpenFPGA, i.e., an open-source eFPGA generator, precisely because we can explore different parameters and produce the corresponding fabrics -- specifically for redaction -- that are smaller than commercial eFPGAs. 
In fact, commercial eFPGA fabrics are less flexible, closed, and typically larger as they prioritize other design goals (e.g., FlexLogic fabrics~\cite{flexlogic} start at $\sim$1K LUTs). 
We note as well that redacting parts of an IP with small fabrics can already incurs considerable overheads~\cite{mohan_hardware_2021}.
Next, we will introduce the \ac{eFPGA} architecture, describing which parameters we consider in our analysis.

\section{Background on eFPGA\label{sec:background}}
This section provides a brief overview of FPGAs, covering the most crucial parts, i.e., architectural choices and EDA toolchains for agile hardware development techniques. 
These are the essential factors for enabling eFPGA redaction, as explored in this paper. 
We refer the readers to the work of Boutros and Betz for more details on FPGA architectures~\cite{FPGA_arch}. 

\subsection{FPGA Architectures\label{sec:fpga_arch}}
FPGAs are reconfigurable fabrics that are (re)-programmable ``in the field'' to implement a specific digital design. 
Modern FPGAs are designed using a tile-based architecture, where the FPGA comprises repeatable tiles and a ``sea'' of routing resources, as shown in \autoref{fig:fpga}~(\ding{202}). 
A \textit{B$\times$B} architecture means there are \textit{B} tiles distributed in horizontal and vertical direction, respectively. For example, \autoref{fig:fpga} (\ding{202}) shows a 5$\times$5 FPGA architecture.
The predominant tiles in an FPGA are \ac{CLB} tiles that implement logic functions. 
An example of a \ac{CLB} tile is shown in \autoref{fig:fpga} (\ding{203}); it contains a \ac{CLB} and blocks for setting the connection between signals within and outside the tile. 
Modern FPGAs can also include some specialized tiles, such as block RAM (BRAM) or digital signal processing (DSP) tiles. 
A heterogeneous tile-based FPGA gives the designer flexibility to meet design requirements and also control the \ac{PPA} aspects of the architecture.
Tile-based architectures offer a better trade-off between programmability and efficiency compared to alternatives ~\cite{FPGA_arch}; designers can also separately focus on the problem of how to route and connect signals within a tile, and problem of interconnecting tiles ``globally''. 
This allows engineers to focus on optimizing the layout of a tile and spend less time on placing and routing tiles. 
At a lower level of abstraction, the building blocks of an FPGA include the following: 

\begin{figure}[t]
\centering
\includegraphics[width=1\columnwidth]{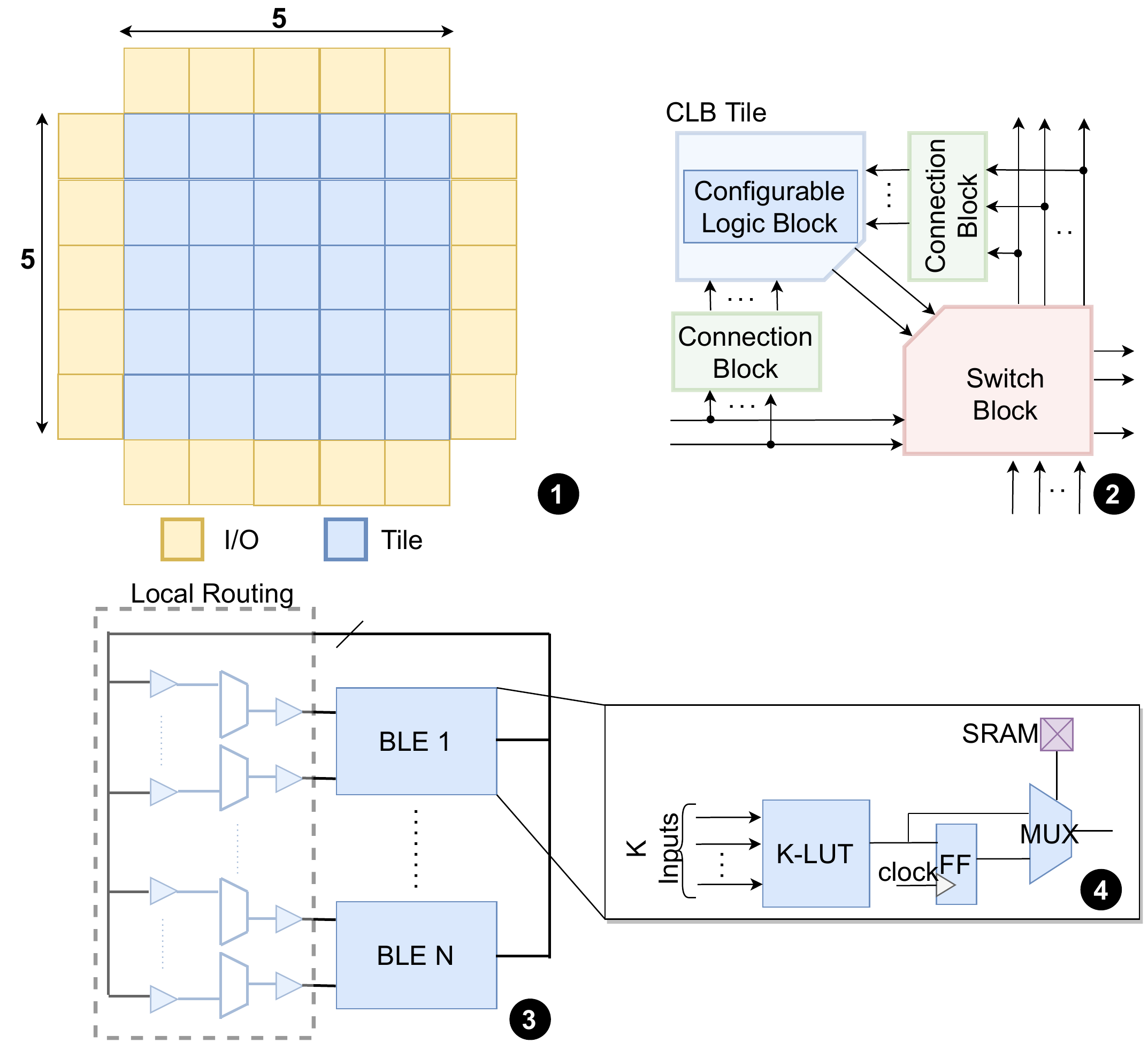}
\caption{Simplified view of an FPGA and its constituent parts.  
}
\label{fig:fpga}
\end{figure}

\textbf{Configurable Logic Blocks} are used to implement combinational and sequential logic. 
\autoref{fig:fpga} (\ding{204}) shows a detailed \ac{CLB} architecture, where there are $N$ \textbf{\acp{BLE}} which are connected through a \textit{local routing architecture}. 
A \ac{BLE} is the primitive module implementing logic functions and comprises a \ac{LUT}, a \ac{FF} and a 2-input multiplexer, as shown in \autoref{fig:fpga} (\ding{205}). 
One can map a K-input single-output Boolean function to a single K-input LUT. 
By configuring 2-input multiplexer, a \ac{BLE} can operate in either combinational or sequential mode. 
To route interconnect \ac{CLB} inputs and \ac{BLE} inputs and outputs, the local routing architecture, typically implemented as a crossbar, includes a set of programmable multiplexers. 
The local routing guarantees that \acp{BLE} can be fully connected to each other and also to every \ac{CLB} input pin. 
The logic capacity of a \ac{CLB} is determined by the following parameters: 
(1) input size of LUTs, $K$; (2) numbers of \acp{BLE} in a \ac{CLB}, $N$; and (3) number of inputs to the \ac{CLB}, $I$. 
These parameters are chosen based on the trade-off between the logic capacity and impact on the area, delay and power. 
To have better resource utilization in a \ac{CLB}, for any LUT size, $I = \frac{K(N+1)}{2}$ has been shown to give good \ac{PPA}~\cite{LUT}. 

\begin{figure}[t]
     \centering
     \includegraphics[width=1\columnwidth]{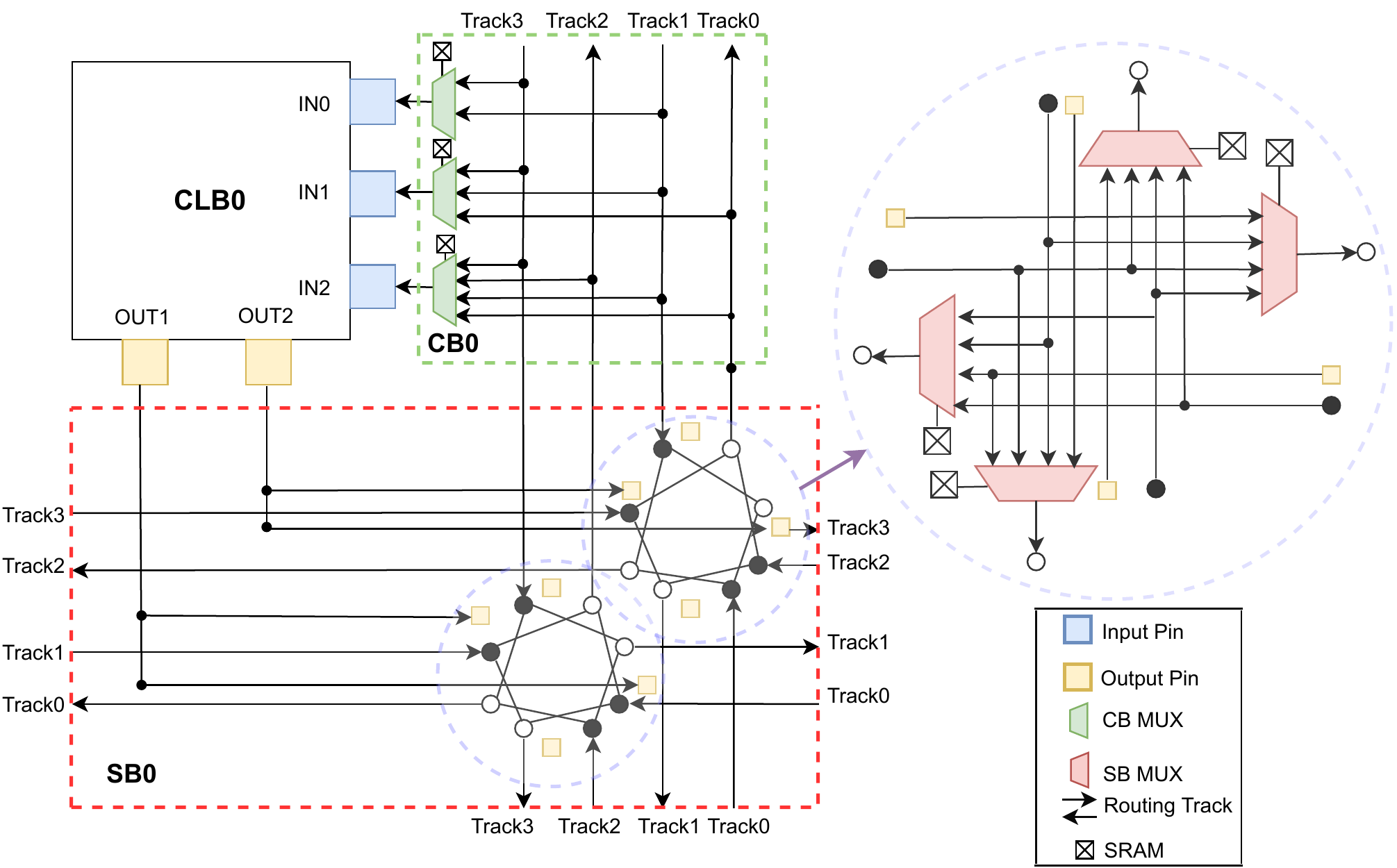}
     \caption{Global Routing Structure.  
    }
     \label{fig:route}
 \end{figure}

The \textbf{Global Routing Architecture} determines the signal routing outside \acp{CLB}, and comprises of \acp{CB} and \acp{SB}. 
Both \acp{CB} and \acp{SB} employ programmable multiplexers for routing. 
\acp{CB} are used to connect the input and output of \acp{CLB} to routing tracks (that connect different tiles) and \acp{SB} connect routing blocks together for producing longer routes between tiles~\cite{FPGA_arch}. 
Typically, a sparse connection is used for global routing where a routing multiplexer is connected to a subset of routing tracks to have a better trade-off between routing area and routability. 
Parameters for routing include: (1) the number of routing tracks grouped together in a channel, $W$; (2) the fraction of routing track connected to a \ac{CLB} input, $F_{c,in}$; (3) the fraction of routing track connected to a \ac{CLB} output, $F_{c,out}$; and (4) the number of routing tracks that can be connected to one routing track, $F_s$.
In modern FPGAs, uni-direction global routing is preferred over classical bi-directional routing~\cite{global_routing}, as it can save 25\% area and improve delay by 9\%. 
\autoref{fig:route} illustrates an example of a unidirectional global routing architecture, where \ac{CLB} \textit{CLB0} is surrounded by a \ac{SB} \textit{SB0} and \ac{CB} \textit{CB0}, with a channel width ($W$) of 4. 
\textit{$F_{c,in}$} of inputs pins \textit{IN0},  \textit{IN1}, and \textit{IN2} are 2/4 = 0.5, 3/4 = 0.75 and 4/4 = 1 respectively. 
\textit{$F_{c,out}$} of output pins \textit{OUT1} and \textit{OUT2} have the same value of 2/4 = 0.5. 
Each routing track connects to 3 other tracks, thus $F_s$, = 3 in \textit{SB0}. 
Usually, a routing path starts from a \ac{CLB} input, and connect to routing track through a \ac{CB}, and then passes through \ac{SB}, to finally reach a \ac{CLB} output through another \ac{CB}. 
But, if \acp{CLB} are far from each other, the routing may have to go through a number of SBs, which will eventually increase the delay. 
To tackle this problem, routing tracks are allowed to span multiple \acp{CLB}. This parameter is defined as the length of routing track $L$, i.e., the number of \acp{CLB} spanned by a routing track.

The FPGA is configured by loading a \textit{bitstream}, where each bit sets some element of the fabric, such as routing configurations and \ac{LUT} contents. One can load the bitstream with frame-based~\cite{koch_fabulous_2021} and scan-chain based~\cite{mohan_top-down_2021,tang_openfpga_2019} configurations. 
For the purpose of our study, we focus on scan-chain based bitstream programming, where the whole bitstream is loaded sequentially, one bit per cycle, with a dedicated clock (\texttt{prog\_clk}) for this purpose.


\subsection{Open Source (e)FPGA Design Flows}
Heterogeneous computing have renewed interest in \acfp{eFPGA} fabrics due to their flexibility and adaptability. 
Commercially, \acp{FPGA} are coupled tightly to processors in a single-chip so that they can act as a programmable accelerator or co-processor \cite{intel_xeon_fpga, PSchiavone_tvlsi_2021}, with benefits like increasing the peak performance of a \ac{SoC} by 3.4$\times$ along with a 2.9$\times$ power reduction.
Different \acp{SoC} can be customized with different \ac{eFPGA} fabrics to best serve a specific application's requirements, e.g., \acp{eFPGA} for machine learning require a high density of \textit{Digital Signal Processing} (DSP) blocks, embedded memories, and other arithmetic accelerators (such as for multiply and accumulate operations). 

\begin{figure}[t]
    \centering
    \includegraphics[width=0.9\columnwidth]{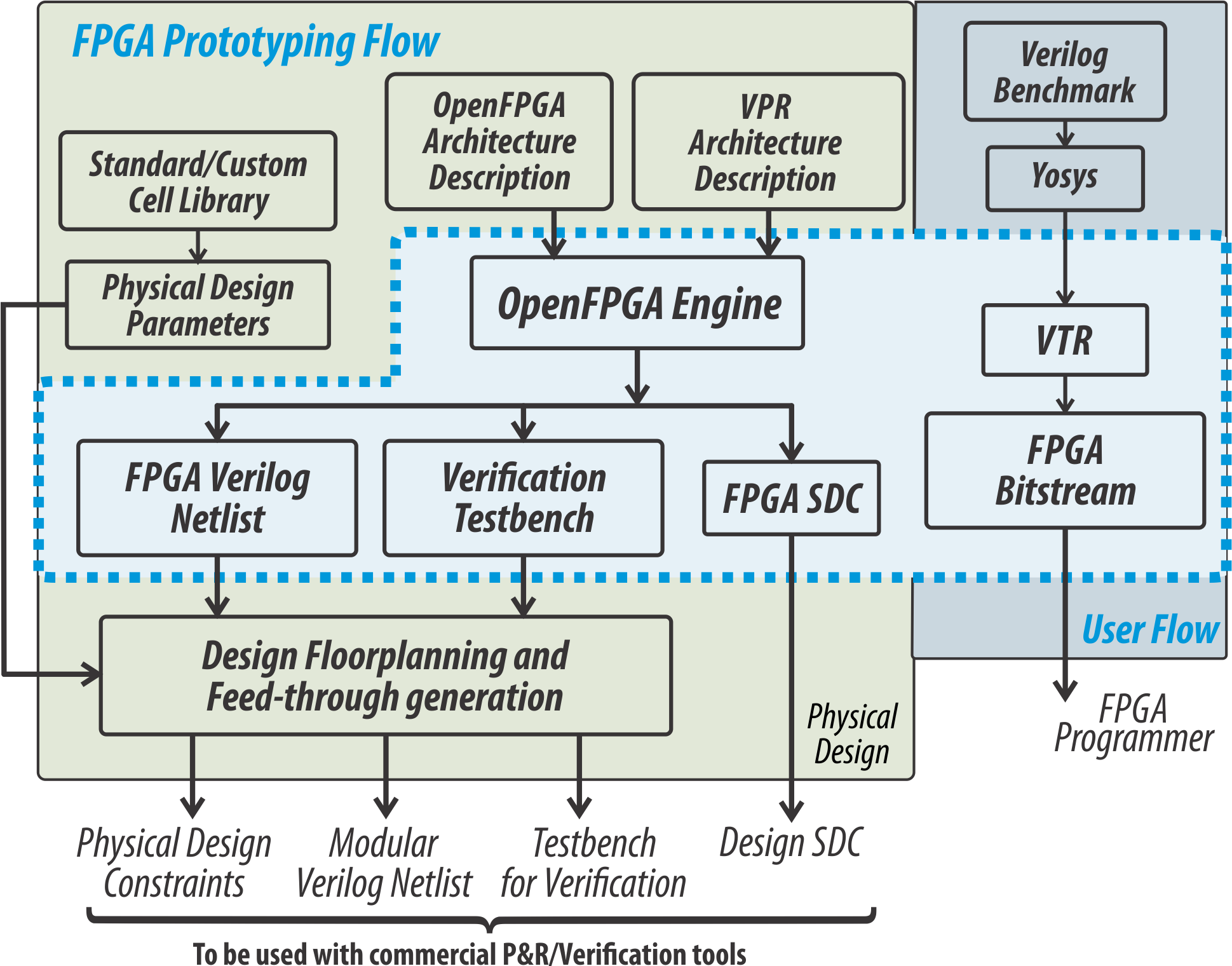}
    \caption{Open-source \ac{eFPGA} design flows: (a) XML-to-layout generation for IC designers; and (b) Verilog-to-Bitstream generation for end-users.}
    \label{fig:openfpga_flow}
\end{figure}

Recently, open-source (e)FPGA prototyping tools have emerged 
\cite{tang_openfpga_2020, koch_fabulous_2021, ALi_FPGA_2021}. 
\autoref{fig:openfpga_flow} illustrates principles of the OpenFPGA framework for prototyping customizable \acp{eFPGA}~\cite{tang_openfpga_2020}.
The framework provides a unified environment for FPGA IP Prototyping and supporting FPGA CAD tools. In the \textit{XML-to-layout flow}, designers produce fabrication-ready \ac{eFPGA} layouts by specifying designs with XML-based architecture description languages \cite{JLuu_FPGA_2011, XTang_tvlsi_2018}, customizing circuit elements, standard cells, and flexible hardware IPs. 
The core engine converts the architecture description Verilog netlists (either tech-mapped or synthesizable). 
The auto-generated netlists are post-processed \cite{GGore_ispd_2021}, with a focus on easing the physical design flow, and then followed by \textit{Place\&Route} (P\&R) tools for generating GDSII layouts and design sign-offs. 
For functional verification, OpenFPGA also produces Verilog testbenches. 
The testbenches validate the correctness of a generated fabric, simulating a complete process, including bitstream downloading and \ac{eFPGA} operation. 
The ability to create custom fabrics is a better fit for redaction compared with off-the-shelf (commercial) \ac{eFPGA} \ac{IP}~\cite{mohan_hardware_2021}.
In the \textit{Verilog-to-Bitstream flow}, end-users can implement HDL designs on the \acp{eFPGA}. HDL designs are first synthesized by Yosys \cite{CWolf_yosys_2013} and physically mapped (packed, placed, and routed) on the \ac{eFPGA} programmable resources using VPR \cite{KMurray_trets_2020} tool. The implemented design is translated to a bitstream which is compatible with configuration protocols of \acp{eFPGA}.

Open-source efforts aim to overcome two major technical barriers of contemporary \ac{eFPGA} development: (1) the time-consuming physical design process---by leveraging the sophisticated ASIC design tools rather than manual layouts, and (2) the increasing design complexity of associated \ac{EDA} tool-chain---by using well-known open-source \ac{FPGA} architecture exploration tools, e.g., VPR~\cite{KMurray_trets_2020}, rather than developing ad hoc, in-house tools.
Using the design flows in \autoref{fig:openfpga_flow}, the development cycle of a 160k-\ac{LUT} \ac{FPGA} layout is $\sim$24~hours and its performance is competitive against commercial products \cite{GGore_ispd_2021, tang_openfpga_2020}. 
We thus adopt the OpenFPGA framework to implement \ac{eFPGA} fabrics \cite{tang_openfpga_2020} for redaction and provide insights into our experience.

\subsection{eFPGA Architecture for this Study\label{sec:our-fabrics}} 
Given the different FPGA architectural parameters, as we described in \autoref{sec:fpga_arch}, we are interested in understanding how changing the parameters affects security. 
As the design space is considerable, we limit our study to exploring the $K$ and $N$ parameters, as these are the factors that have direct impact on complexity and hardware utilization.
In the FPGA fabric netlists, LUTs comprise a tree-of-MUXes, so varying N and K changes the MUX sizes.

To summarize, \autoref{tab:arch_param} describes the architecture settings that we explore for generating eFPGA redaction fabrics. 
For this study, we use a tile-based FPGA to emulate complexity similar to that in a commercial FPGA, and vary the overall fabric size from 4$\times$4 to 6$\times$6 tiles. 
For consistency of results, we use uni-directional routing. 
These parameters are set on the basis of the prior studies of FPGA designs~\cite{LUT, FPGA_arch}, which showed that these ranges of parameters give the best power, performance, and area (PPA) results. 
In this work, we first characterize the fabrics' bitstream size, area, power, and delay, and then evaluate their security in \autoref{sec:security-I} and \autoref{sec:security-II}. 

\begin{table*}[h]
    \centering
    \caption{eFPGA Fabric Architecture Parameters used in this work. 
    }
      \label{tab:arch_param}
   \begin{tabular}{ c c c}
   \toprule
Parameter & Range & Description \\ 
\midrule
K & [3,7] & Input size of a LUT \\ 
N & [2,8] & Number of \acp{BLE} in a Configuration Logic Block \\ 
I & $I = \frac{K(N+1)}{2}$ & Number of inputs to a CLB. \\ 
W & 40 &  Number of routing tracks in a channel. \\ 
$F_{c,in}$ & 0.15 & Fraction of routing tracks to which each \ac{CLB} input pin connects \\ 
${F_{c,out}}$ & 0.1 & Fraction of routing tracks to which each \ac{CLB} output pin connects. \\ 
$F_s$ & 3 & Number of routing tracks to which each incoming routing track
can connect in a \ac{SB}. \\ 
L & 4 & The length of a routing track in term of the number of \acp{CLB} spanned
by the track \\ \bottomrule
\end{tabular}
  
\end{table*}


\subsubsection{Tool Setup for Fabric Design} 
To generate the fabrics and determine the cost of the redaction fabrics based on area, power, and delay, we synthesize, place, and floorplan fabrics that we generate using the OpenFPGA flow~\cite{XTang_tvlsi_2018}. 
This flow is depicted in \autoref{fig:redact-flow}. 
For synthesis we use Cadence Genus 18.14 and for layout implementation we use Cadence Innovus 18.10. 
The timing, area, and power reports are generated by Innovus. 
For floorplanning we set utilization to 70\% for faster timing closure. We use FreePDK 45~nm library~\cite{FreePDK45} for our study. 

\subsubsection{Bitstream Characteristics}
\begin{table*}[h]
\centering
\caption{Bitstream size for different eFPGA fabrics for different configuration parameters.}
\label{tab:bitstream_result}
\resizebox{\textwidth}{!}{
\begin{tabular}{c ccccc ccccc ccccc}
\toprule
\multirow{2}{*}{\diagbox[]{N}{K}} & \multicolumn{5}{c}{Fabric: 4$\times$4} & \multicolumn{5}{c}{Fabric: 5$\times$5} & \multicolumn{5}{c}{Fabric: 6$\times$6} \\ 
\cmidrule(l){2-6} \cmidrule(l){7-11} \cmidrule(l){12-16} 
  & { 3 } & { 4 } & { 5 } & { 6 } & { 7 } & { 3 } & { 4 } & { 5 } & { 6 } & { 7 } & { 3 } & { 4 } & { 5 } & { 6 } & { 7 } \\ \midrule
2 & 642 & 680 & 856 & 1166 & 1712 & 1265 & 1553 & 1958 & 2684 & 3893 & 2090 & 2739 & 3322 & 5037 & 7313 \\
3 & 766 & 851 & 1107 & 1559 & 2475 & 1539 & 1908 & 2493 & 3501 & 5725 & 2574 & 3309 & 4430 & 6635 & 10326 \\
4 & 878 & 1090 & 1518 & 2019 & 3262 & 1786 & 2370 & 3344 & 4974 & 7150 & 3533 & 4392 & 6137 & 8598 & 13144 \\
5 & 982 & 1357 & 1782 & 2775 & 3958 & 2354 & 3161 & 3926 & 6748 & 8710 & 3935 & 5369 & 6786 & 11510 & 15525 \\
6 & 1195 & 1509 & 2053 & 2969 & 4661 & 2429 & 3302 & 4780 & 6948 & 10693 & 4679 & 6059 & 8251 & 11947 & 18348 \\
7 & 1284 & 1688 & 2312 & 3388 & 5552 & 3030 & 3948 & 5379 & 7598 & 12714 & 5134 & 6766 & 9310 & 13770 & 21660 \\
8 & 1434 & 1890 & 2614 & 4030 & 6361 & 3321 & 4356 & 6007 & 9211 & 14378 & 5647 & 7487 & 10420 & 16116 & 25305 \\ \bottomrule
\end{tabular}}
\end{table*}

\noindent \autoref{tab:bitstream_result} reports the number of bits required for the configuration bitstream, and this can be taken as a measure of the overall ``programmability'' of the fabric. 
It is important to note that there are overlaps in bitstream sizes across tile sizes (e.g., 4$\times$4 K7N8 fabric has more bits in its bitstream compared to  5$\times$5 K5N7). 

\subsubsection{Area Characteristics}
\begin{table*}[h]
\centering
\caption{Area ($mm^2$) for different eFPGA fabrics.}
\label{tab:area_result}
\resizebox{\textwidth}{!}{
\begin{tabular}{c ccccc ccccc ccccc}
\toprule
\multirow{2}{*}{\diagbox[]{N}{K}} & \multicolumn{5}{c}{Fabric: 4$\times$4} & \multicolumn{5}{c}{Fabric: 5$\times$5} & \multicolumn{5}{c}{Fabric: 6$\times$6} \\ 
\cmidrule(l){2-6} \cmidrule(l){7-11} \cmidrule(l){12-16} 
  & { 3 } & { 4 } & { 5 } & { 6 } & { 7 } & { 3 } & { 4 } & { 5 } & { 6 } & { 7 } & { 3 } & { 4 } & { 5 } & { 6 } & { 7 } \\ \midrule

2 & 0.006	&0.007	&0.008&	0.011&	0.016& 0.012	&0.016	&0.019	&0.025	&0.037 &0.020&	0.028&	0.034&	0.051&	0.070\\ 
3 & 0.007&	0.008&	0.011&	0.016&	0.023 &0.015	&0.018	&0.026	&0.033	&0.053 &0.025&	0.032&	0.047&	0.066&	0.096\\
4 & 0.009&	0.011&	0.016&	0.024&	0.031 &0.018	&0.024	&0.035	&0.049	&0.069 &0.036&	0.046&	0.068&	0.087&	0.127 \\ 
5 & 0.010&	0.014&	0.018&	0.029&	0.040 &0.025	&0.033	&0.040	&0.069	&0.085 &0.042&	0.056&	0.067&	0.122&	0.159\\ 
6 & 0.013&	0.016&	0.022&	0.032&	0.046 &0.027	&0.035	&0.052	&0.077	&0.107  &0.053&	0.065&	0.086&	0.128&	0.183 \\ 
7 & 0.014&	0.019&	0.026&	0.038&	0.058 &0.034	&0.043	&0.059	&0.087	&0.133 & 0.058&	0.073&	0.101&	0.150&	0.220\\ 
8 & 0.017&	0.023&	0.031&	0.047&	0.068 &0.039	&0.050	&0.068	&0.104	&0.155 & 0.067&	0.086&	0.117&	0.156&	0.224\\ \bottomrule
\end{tabular} } 
\end{table*}

\noindent \autoref{tab:area_result} shows how the area is affected by varying $N$ and $K$. 
For a given $N$ value, as we increase $K$ the number of inputs to the \ac{CLB} increases (as a result of the relationship between $I$, $K$, and $N$ mentioned in \autoref{sec:fpga_arch}); increasing the LUT sizes in the \acp{BLE} and also slightly affecting local and global routing. 
For a given $K$ however, an increase in $N$ has more impact in increasing the area, as entire \acp{BLE} are added; this increases the complexity of both local and global routing, as suddenly there is a jump of another K inputs to the \acp{CLB}, resulting in added pressure on local routing to route this additional set of inputs to \acp{CLB}. 
This will also force the global routing (\acp{CB} and SBs) to increase the routing complexity as more number of inputs are being fed to CLBs.




\subsubsection{Delay Characteristics}
\begin{table*}[h]
\centering
\caption{Critical Path Delay ($ns$) for different eFPGA fabrics.}
\label{tab:delay_result}
\resizebox{0.9\textwidth}{!}{
\begin{tabular}{c ccccc ccccc ccccc}
\toprule
\multirow{2}{*}{\diagbox[]{N}{K}} & \multicolumn{5}{c}{Fabric: 4$\times$4} & \multicolumn{5}{c}{Fabric: 5$\times$5} & \multicolumn{5}{c}{Fabric: 6$\times$6} \\ 
\cmidrule(l){2-6} \cmidrule(l){7-11} \cmidrule(l){12-16} 
  & { 3 } & { 4 } & { 5 } & { 6 } & { 7 } & { 3 } & { 4 } & { 5 } & { 6 } & { 7 } & { 3 } & { 4 } & { 5 } & { 6 } & { 7 } \\ 
  \midrule
2 & 1.85 & 1.96 & 1.91 & 1.22 & 1.68 & 3.74 & 2.82 & 2.24 & 2.24 & 1.67 & 6.47 & 4.41 & 3.74 & 3.84 & 4.46 \\ 
3 & 1.86 & 1.52 & 1.91 & 1.39 & 1.10 & 3.15 & 2.37 & 3.63 & 3.58 & 2.64 & 5.64 & 3.78 & 5.91 & 3.81 & 4.11 \\
4 & 1.88 & 1.29 & 1.41 & 1.28 & 1.56 & 3.99 & 3.21 & 3.15 & 3.59 & 2.48 & 6.93 & 5.24 & 5.07 & 4.23 & 3.61 \\ 
5 & 1.77 & 2.06 & 1.65 & 1.77 & 1.35 & 5.34 & 3.45 & 4.14 & 3.36 & 3.37 & 8.92 & 6.21 & 4.89 & 6.19 & 3.74 \\ 
6 & 2.58 & 1.67 & 1.65 & 1.88 & 1.41 & 5.30 & 3.39 & 3.93 & 3.21 & 2.52 & 8.72 & 6.21 & 4.86 & 5.28 & 3.51 \\ 
7 & 2.55 & 2.43 & 1.83 & 1.59 & 1.23 & 6.43 & 4.83 & 3.99 & 3.21 & 2.58 & 8.73 & 5.89 & 4.74 & 5.22 & 3.65 \\ 
8 & 2.43 & 2.12 & 1.77 & 1.82 & 1.83 & 6.52 & 4.71 & 3.72 & 3.06 & 2.81 & 8.74 & 5.94 & 4.54 & 5.02 & 3.33 \\ \bottomrule
\end{tabular} 
} 
\end{table*}

\noindent Compared to our study on area, the impact on the critical path delay from varying $N$ and $K$ is less obvious, as shown in \autoref{tab:delay_result}. 
As observed in prior work~\cite{LUT, FPGA_arch}, the impact on delay is not a linear function of $K$ and $N$. 
In our fabrics, we observe that for a given $N$, the delay values improve as one increases $K$, where the least delay is generally achieved for the largest $K$ ($=7$). 

\subsubsection{Power Characteristics}
\begin{table*}[h]
\centering
\caption{Power ($mW$) for different eFPGA fabrics.}
\label{tab:power_result}
\resizebox{\textwidth}{!}{
\begin{tabular}{c ccccc ccccc ccccc}
\toprule
\multirow{2}{*}{\diagbox[]{N}{K}} & \multicolumn{5}{c}{Fabric: 4$\times$4} & \multicolumn{5}{c}{Fabric: 5$\times$5} & \multicolumn{5}{c}{Fabric: 6$\times$6} \\ 
\cmidrule(l){2-6} \cmidrule(l){7-11} \cmidrule(l){12-16} 
  & { 3 } & { 4 } & { 5 } & { 6 } & { 7 } & { 3 } & { 4 } & { 5 } & { 6 } & { 7 } & { 3 } & { 4 } & { 5 } & { 6 } & { 7 } \\ \midrule
2 & 0.71 & 0.79 & 0.98 & 1.31 & 1.94 & 1.56 & 2.06 & 2.45 & 3.19 & 4.85 & 2.62 & 3.75 & 4.45 & 6.76 & 9.24 \\
3 & 0.89 & 0.98 & 1.39 & 1.92 & 2.84 & 1.88 & 2.36 & 3.33 & 4.25 & 6.98 & 3.22 & 4.18 & 6.44 & 8.64 & 13.06 \\ 
4 & 1.07 & 1.39 & 1.99 & 2.63 & 3.87 & 2.33 & 3.34 & 4.68 & 6.35 & 9.14 & 4.84 & 6.34 & 9.31 & 11.63 & 17.23 \\ 
5 & 1.31 & 1.77 & 2.35 & 3.75 & 5.03 & 3.30 & 4.52 & 5.46 & 9.35 & 11.16 & 5.74 & 7.92 & 9.48 & 16.57 & 21.44 \\ 
6 & 1.62 & 2.09 & 3.02 & 4.02 & 5.58 & 3.53 & 4.86 & 7.44 & 10.35 & 13.68 & 7.36 & 9.44 & 12.61 & 17.56 & 23.13 \\ 
7 & 1.92 & 2.51 & 3.58 & 4.86 & 7.14 & 4.89 & 6.19 & 8.69 & 11.41 & 16.12 & 8.46 & 10.80 & 15.12 & 20.63 & 28.31 \\ 
8 & 2.20 & 3.07 & 4.28 & 5.97 & 8.17 & 5.65 & 7.33 & 10.21 & 14.14 & 19.52 & 9.73 & 12.98 & 17.95 & 21.44 & 30.43 \\ \bottomrule
\end{tabular}}
\end{table*}

\noindent Power is shown in \autoref{tab:power_result}. 
Similar to what we observed from area, power increases with increasing complexity of the fabric (increasing $K$ and $N$). 
If one look at the two extremes for a given fabric size, \{$K$=3 and $N$=2\} and the other being \{$K$=7 and $N$=8\}, there is almost $10\times$ increase in power consumption. 


\subsection{General Observations\label{sec:question}}
There is considerable variation in bitstream size, area, delay, and power as we vary $K$ and $N$, given a fabric size. 
Given a module to redact, a designer will naturally be drawn to the fabric configuration with the least area/power/delay that can fit the redaction target. 
However, let us consider bitstream size as our security parameter (intuition: more bits in the bitstream, more security). 
Seeing as there are fabrics that have similar configuration bitstream sizes with different fabric sizes and $K$/$N$ parameter values, this begs the question: \textbf{Can we gauge security by considering only the bitstream size?}
Take, for instance, the 4$\times$4 K7N8 fabric uses 6361 bits for its bitstream, does this mean better security compared to the 5$\times$5 K5N7 fabric, which $\sim$1000 fewer bits \textit{and} requires smaller area? 
In the next section, we perform a security analysis on all the fabrics to try to see if this is indeed the case.

\section{Assessing eFPGA-based Redaction Fabrics\label{sec:security-I}}
This section describes the threat model and assumptions under which our study operates, outlines our intuitions about the characteristics of eFPGAs generally that contribute to their security, and then present the results of our experiments. 
We perform all our experiments using a High Performance Computing (HPC) environment, with jobs running in parallel, each of them on an independent compute node that has an Intel Xeon Platinum 8268 processor running at 2.9~GHz and 256~GB of RAM.

\subsection{Assumptions and Threat Model} 
For insight into the security offered by using \ac{eFPGA}-based redaction, we explore SAT-attack resilience of the various fabrics (as described earlier in \autoref{sec:our-fabrics}), as this has been used to gauge the security of redaction in prior work~\cite{mohan_hardware_2021} and has proven to be a challenge to overcome for prior IP protection approaches~\cite{LLC}. 
Previous work suggests that large \ac{FPGA} bitstream lengths make SAT-based attacks impractical~\cite{hu_functional_2019} and the evaluation results in ~\cite{mohan_hardware_2021} appear to support this claim. 

As we want to investigate how structural variations of the \ac{eFPGA} contribute to complexity parameters of SAT-based attacks, we perform a security evaluation by launching a SAT-based attack on the fabrics described in \autoref{sec:our-fabrics}. 
In our analyses, we assume that the designers already know which parts of the design must be protected to stay competitive in the market. 
Hence, this paper does not address the selection of the modules to be redacted; we assume that a given fabric is already selected as sufficient for their desired redaction. 
For worst-case analysis, our threat model overwhelmingly favors the attacker. 
We assume the attacker has access to the redacted \ac{IC}'s netlist and to a fully-scanned\footnote{For those unfamiliar with Design-for-Test concepts, a fully-scanned design means that all flip-flops have been replaced with scan flip-flops that are connected to each other into a long scan chain, essentially acting as a large shift register. A tester is able to scan-in values and then scan-out the combinational logic outputs a clock cycle later, thus deducing the input/output relationship for the logic in the design.} and fully-unlocked design (i.e., access to an Oracle with the bitstream loaded). 

The adversary has to override three challenges before they can launch a SAT-based attack on \ac{eFPGA} fabric. 
First, they have to isolate the \ac{eFPGA} fabric from the rest of the \ac{IP}; this is possible since the regular structure of the fabric is distinguishable from the rest of the design. 
Second, for the Oracle, the adversary should have complete control over the inputs, outputs and internal flip-flops, excluding configurable flip-flops. 
We endow the attacker with these capabilities although there are orthogonal efforts to mitigate this Oracle-based threat model~\cite{limaye_thwarting_2020}. 
Third, the adversary cannot extract the \ac{FPGA} bitstream~\cite{hu_functional_2019}, i.e., the attacker does \textit{not} have access to the configuration flip-flops, so cannot simply steal the bitstream directly. 
While there are a number of attacks on FPGA security~\cite{trimberger_fpga_2014}, we consider such attacks orthogonal to this study. 
Physical attacks (e.g., optical probing~\cite{rahman_key_2020}) are out of scope. 
Our threat model and assumptions are consistent with prior work~\cite{mohan_hardware_2021}. 

\subsection{Security Evaluation Setup}
In an \ac{eFPGA}, the bitstream is loaded into configuration \acp{FF}.
The configuration \acp{FF} are interconnected as a scan-chain driven by a programming clock (\texttt{prog\_clk}).
To prepare the fabrics that we generated in \autoref{sec:our-fabrics}, we need to transform the gate-level netlist and produce a netlist understood by an attack tool, with the configuration bitstream as a set of ``key inputs''. 
To identify the configuration scan chain, we do a depth-first search of the netlist, starting from the \texttt{scan\_in\_head} port, until we reach the \texttt{scan\_in\_tail}. 
All \acp{FF} in the traversal path driven by the programming clock (\texttt{prog\_clk}) store the configuration bitstream. 
The order in which the configuration \acp{FF} are detected corresponds to the bitstream order. 
The detected configuration \acp{FF} are exposed as primary key inputs to convert the \ac{eFPGA} netlist into a netlist suitable for SAT attack. 
This netlist is fed to IcySAT~\cite{shamsi_icysat_2019} to unroll hard loops (as will be explained next). 
To model an Oracle, we use the same locked netlist, but set the key bits to the configuration values from the bitstream generated in the OpenFPGA flow. 
The unrolled netlist and the oracle netlist are used with the \texttt{KC2} attack tool~\cite{shamsi_kc2_2019}. 

\subsection{On Combinational Loops in eFPGAs}
The SAT-based attack requires an attacker to model a miter circuit featuring the design-under-attack as input to a SAT solver~\cite{subramanyan_evaluating_2015}; for an eFPGA fabric, the configuration bitstream is the ``key''. 
There are several factors that make a SAT solver's task challenging. 
SAT solvers fail in the presence of combinational loops~\cite{cyc_sat}, as these lead to unstable results or repeated distinguishing input patterns. 
Note that, in well-formed designs, circuits with structural combinational loops are usually designed such that the overall design behaves as though it is acyclic. 
The structure of \acp{eFPGA} includes several instances of such loops due to the re-configurable routing network in the fabric. 
The sequence of re-configurable logic represented by the chain of \acp{LUT}/\acp{CLB} interconnected by this re-configurable network adds a polynomial complexity to a SAT formulation. 

To launch a SAT attack on designs with loops, like in eFPGAs, one needs to preprocess the netlist to break the loops and create an acyclic equivalent. 
Researchers have proposed multiple approaches to modify the SAT attack for cyclic designs~\cite{cyc_sat,be_sat,shamsi_icysat_2019}. 
We observe that \acp{eFPGA} have hard combinational loops that CycSAT~\cite{cyc_sat} cannot resolve. 
These hard loops are intertwined such that, when CycSAT breaks a loop to make the circuit acyclic, at least one loop remains. 
The acyclic constraints generated by CycSAT overlook such loops and live-locks the solver into repeating the same \acp{DIP}. 
Be-SAT~\cite{be_sat} can break such loops by pruning the keys leading to live-lock \acp{DIP}. However, it has exponential complexity in key size. 
IcySAT\footnote{More specifically, IcySAT-II, which we will simply refer to as IcySAT in this paper}~\cite{shamsi_icysat_2019} is a loop-breaking alternative that finds a subset of feedback nets that, when ``removed'', make the netlist acyclic. 
The circuit is ``unrolled'' with respect to these feedback nets, with an unroll factor equal to the size of the feedback set. 
Unrolling involves replicating a circuit several times and connecting the feedback wires (that caused the loops) across replicas to represent different time frames (refer to Shamsi {\em et al.}'s work for the definitive explanation~\cite{shamsi_icysat_2019}). 
The unrolled circuit can be fed into a SAT attack tool.

\subsection{Relating Attack Complexity to eFPGA Parameters}
There are several ways to thwart an attacker (or at least, their SAT solver). 
One is to make the circuit very large, such that its representation as a Boolean formula requires an impractical amount of memory to load for the solver -- the need to ``unroll'' loops increases security in this way, as formula sizes grow due to the need for circuit replication or additional constraining clauses.  
In fact, the time complexity of the IcySAT attack that we use is directly related to the total number of clauses and variables that needs to be solved by the solver in retrieving the eFPGA bitstream. 
This complexity can be directly related to the constraints added in a single SAT attack iteration, which is proportional to the gate size of fabric-under-attack. 
Since IcySAT requires unrolling as part of preprocessing to eradicate combinational cycles, the net gate size of the unrolled netlist is proportional to the unroll factor. 
eFPGA fabrics are naturally loop-ridden, arising from the sophisticated intra-CLB (local routing) and inter-CLB (global routing) routing networks. 
Also, another source of variables in the formula is the presence of \acp{LUT} within the \acp{CLB}, where the contents of \ac{LUT} should be determined to reverse-engineer the logic functionality. 
Prior work \cite{LUT_LOCK} explains how programmable logic renders SAT complexity. 
A SAT solver might encounter difficulties due to the polynomial complexity in solving interdependent clauses despite the total number of clauses being very nominal, as is the case with the \acp{LUT} within the eFPGA fabric. 



Recall that, from an architecture perspective, increasing $K$ increases the complexity of local routing within a CLB, while slightly improving the complexity of global routing. Larger \acp{LUT} require larger numbers of crossbar multiplexers to multiplex the LUT fanout and the inputs from connection block. This increases local routing complexity. Meanwhile, increasing $N$ (number of $K$-input \acp{LUT} in a CLB), 
should slightly increase the local routing complexity while significantly increasing the global routing complexity. 

\subsection{Results}

\begin{table}[t!]
\centering
\caption{IcySAT attacks on architectural variants of 4x4 eFPGA fabric. TO represents \textit{time-out}.}
\label{tab:full_icysat_4x4}
\begin{tabular}{@{}L{1cm} @{}C{1.2cm} @{}C{1.3cm} @{}C{1.3cm} @{}C{1.4cm} @{}C{1.6cm} @{}C{1cm}@{}}
\toprule
Fabric & Unroll factor & Bitstream & \#Gates          & Time    & Variables & Clauses \\ \midrule
K3N2   & 64            & 601       & 4227             & 127.6   & 551293    & 1433840 \\ K3N3   & 70            & 725       & 5179             & 283.8   & 737293    & 1922216 \\ K3N4   & 79            & 837       & 6355             & 6998.15 & 1019357   & 2681792 \\ K3N5   & 73            & 941       & 7686             & 14035.9 & 1135927   & 3006984 \\ K3N6   & 105           & 1154      & 9284              & TO      & --        & --      \\
K3N7   & 85            & 1243      & 10823             & TO      & --        & --      \\
K3N8   & 130           & 1393      & 12405             & TO      & --        & --      \\
K4N2   & 55            & 639       & 4053             & 103.54  & 454039    & 1167026 \\ K4N3   & 57            & 810       & 5439             & 896.15  & 629232    & 1642049 \\ K4N4   & 63            & 1049      & 8230              & TO      & --        & --      \\
K4N5   & 113           & 1316      & 10141             & TO      & --        & --      \\
K4N6   & 112           & 1468      & 12352             & TO      & --        & --      \\
K4N7   & 85            & 1647      & 14616             & TO      & --        & --      \\
K4N8   & 132           & 1849      & 17406             & TO      & --        & --      \\
K5N2   & 65            & 815       & 5185             & 267.8   & 685187    & 1773156 \\ K5N3   & 68            & 1066      & 7235             & 25135.2 & 996664    & 2618621 \\ K5N4   & 105           & 1477      & 11274             & TO      & --        & --      \\
K5N5   & 136           & 1741      & 13817             & TO      & --        & --      \\
K5N6   & 104           & 2012      & 17170             & TO      & --        & --      \\
K5N7   & 144           & 2271      & 20836             & TO      & --        & --      \\
K5N8   & 162           & 2573      & 24635             & TO      & --        & --      \\
K6N2   & 69            & 1125      & 6831              & 2033.91 & 955817    & 2481464 \\
K6N3   & 70            & 1518      & 9976              & TO      & --        & --      \\
K6N4   & 93     &  2089        & 14762                & TO      & --    & --  \\
K6N5   & 93            & 2694      & 20357             & TO      & --        & --      \\
K6N6   & 144           & 2928      & 24089             & TO      & --        & --      \\
K6N7   & 143           & 3347      & 28946             & TO      & --        & --      \\
K6N8   & 164           & 3989      & 35073             & TO      & --        & --      \\
K7N2   & 54            & 1671      & 9559              & TO & --    & -- \\
K7N3   &  95           & 2434      & 14700              & TO      & --        & --      \\
K7N4   & 93     &  2089        & 14762                & TO      & --    & --  \\
K7N5   & 84            & 3221      & 21285             & TO      & --        & --      \\
K7N6   & 101           & 3917      & 27075             & TO      & --        & --      \\
K7N7   & 146           & 4620      & 34003             & TO      & --        & --      \\
K7N8   & 151           & 5511      & 41913             & TO      & --        & --      \\
\bottomrule
\end{tabular}%
\end{table}

\begin{figure}[t]
    \centering
    \subfloat[\label{fig:fixed_k}]{
            \includegraphics[width=0.48\columnwidth]{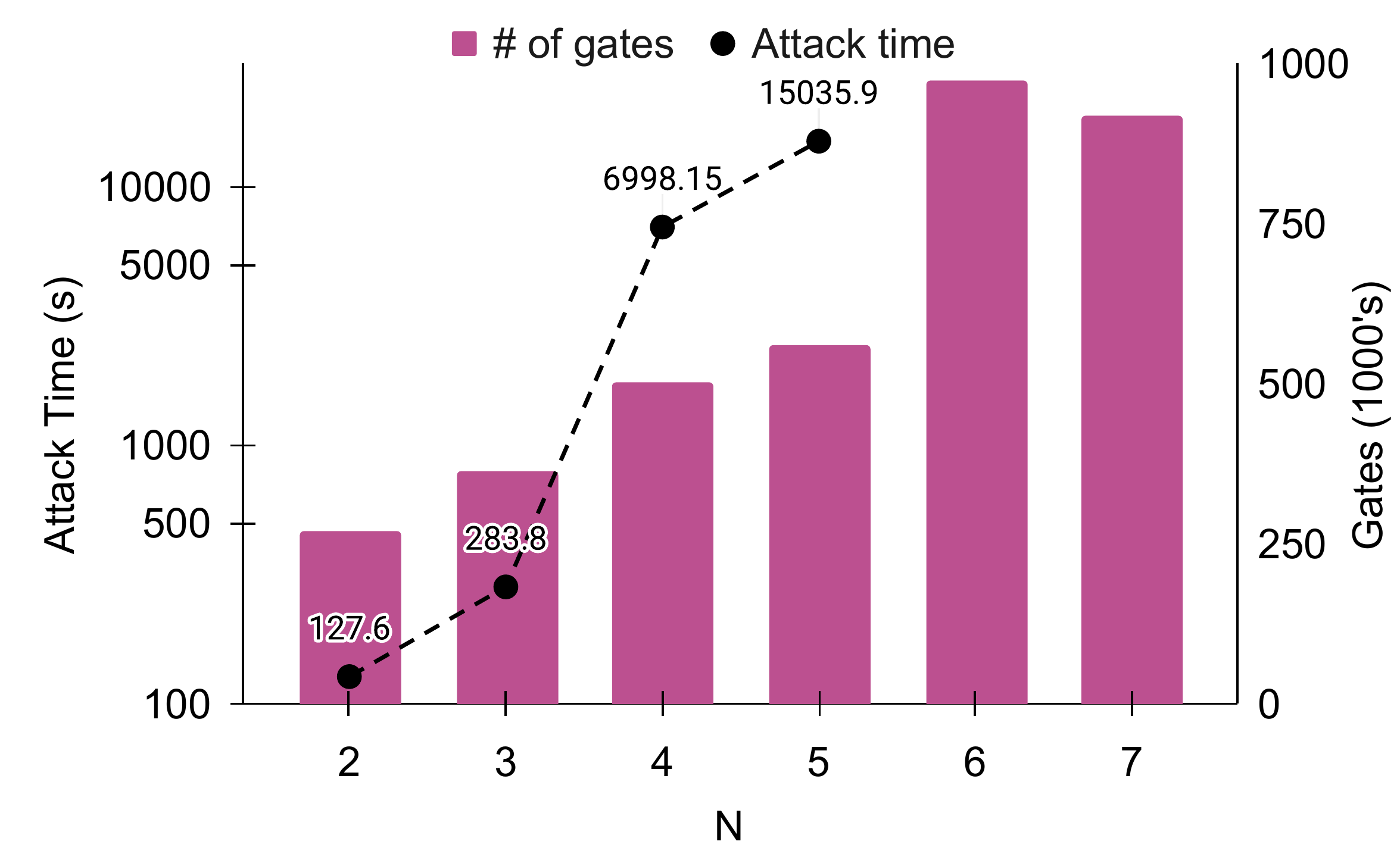}
    }%
    \subfloat[\label{fig:GateN4x4}]{
        {\includegraphics[width=0.48\columnwidth]{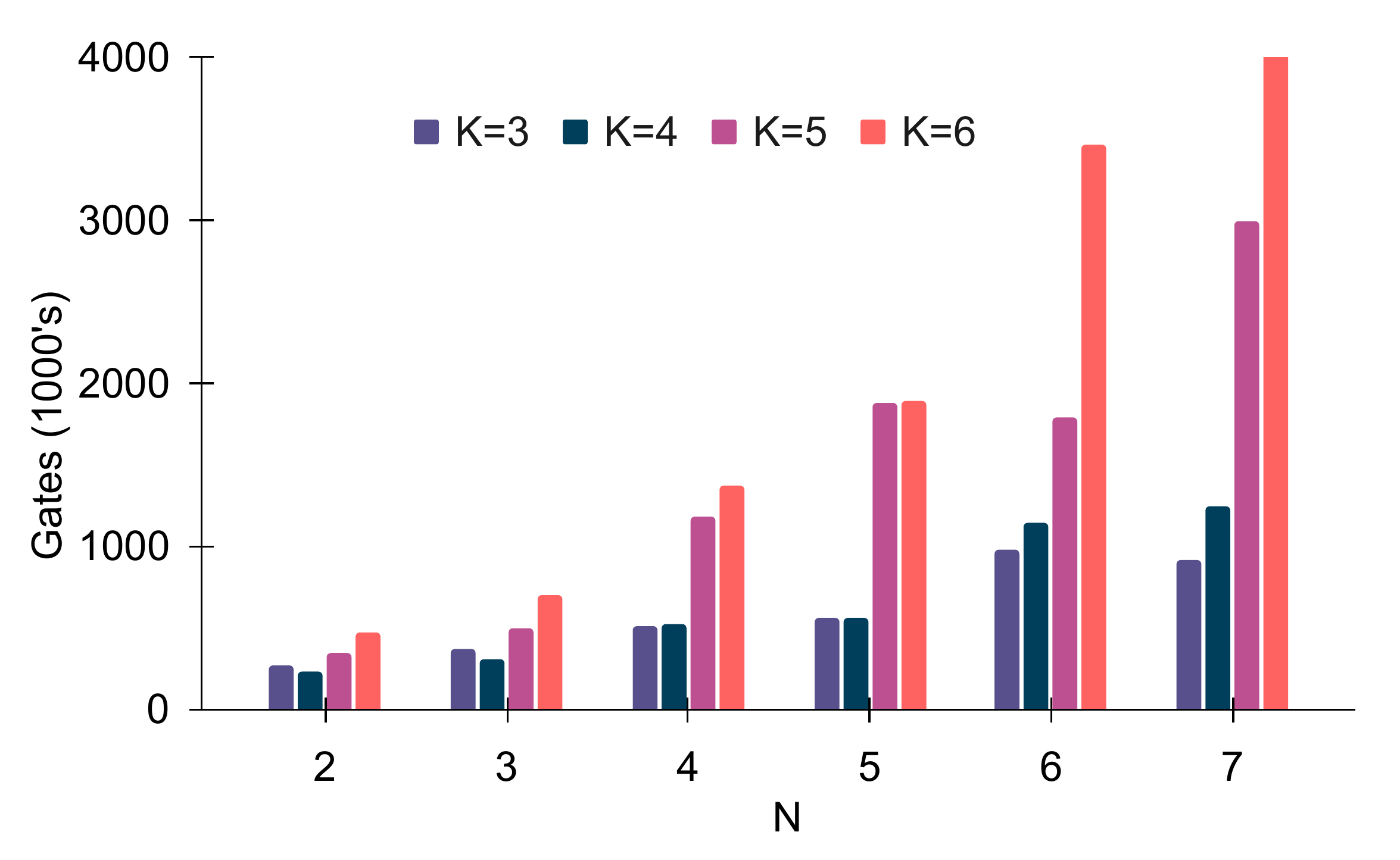}}
    }%
    \caption{(a) Attack time and \# of gates in the unrolled 4$\times$4 fabric for fixed K = 3. The attack timed-out for N = 6, 7. (b) \# of gates in unrolled 4$\times$4 fabric for various K, N.\label{fig:var_n}}
\end{figure}

\begin{figure}[ht]
    \centering
    \subfloat[\label{fig:fixed-N}]{
            \includegraphics[width=0.48\columnwidth]{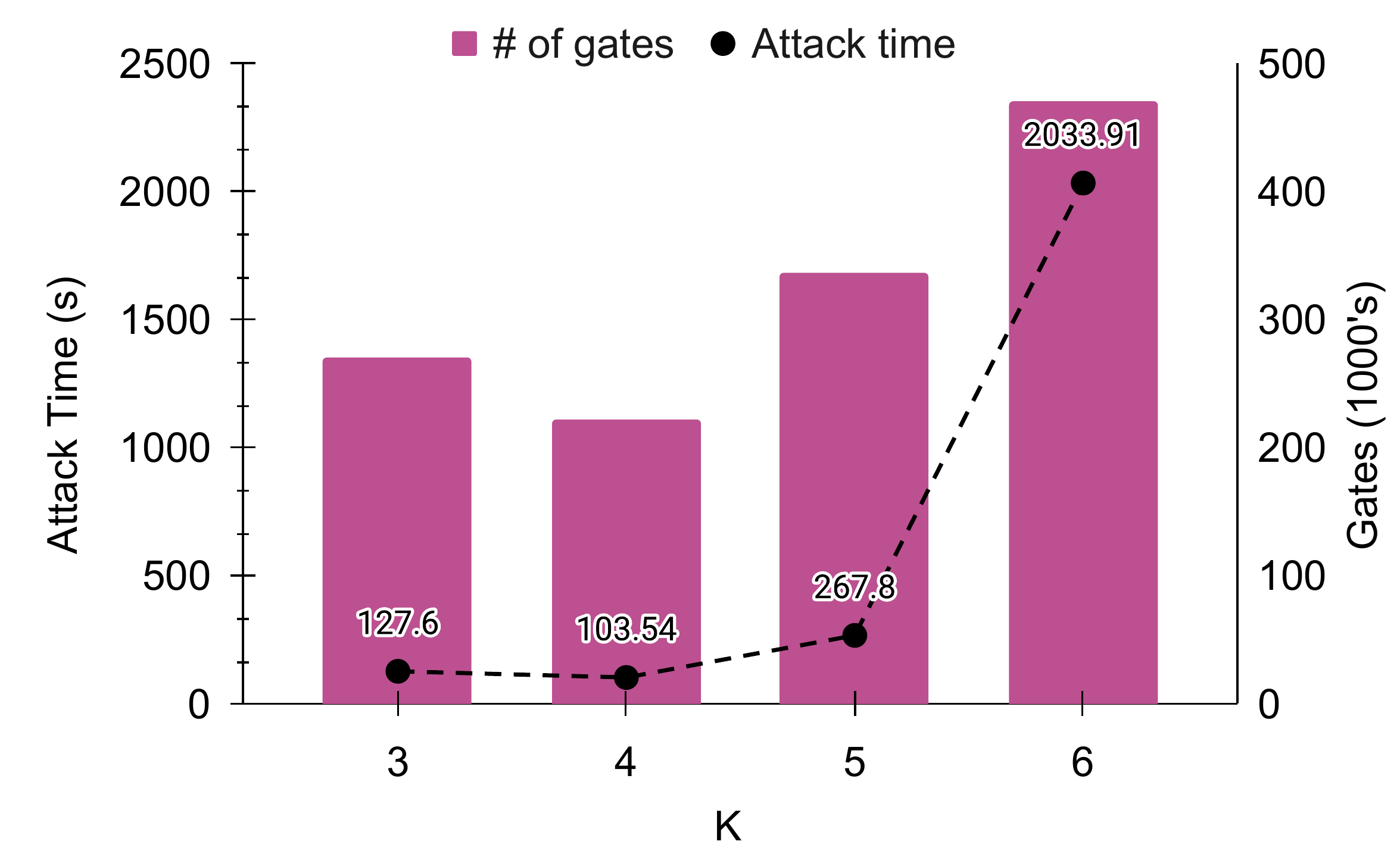}
    }%
    \subfloat[\label{fig:Gate-fixed-N-4x4}]{
        {\includegraphics[width=0.48\columnwidth]{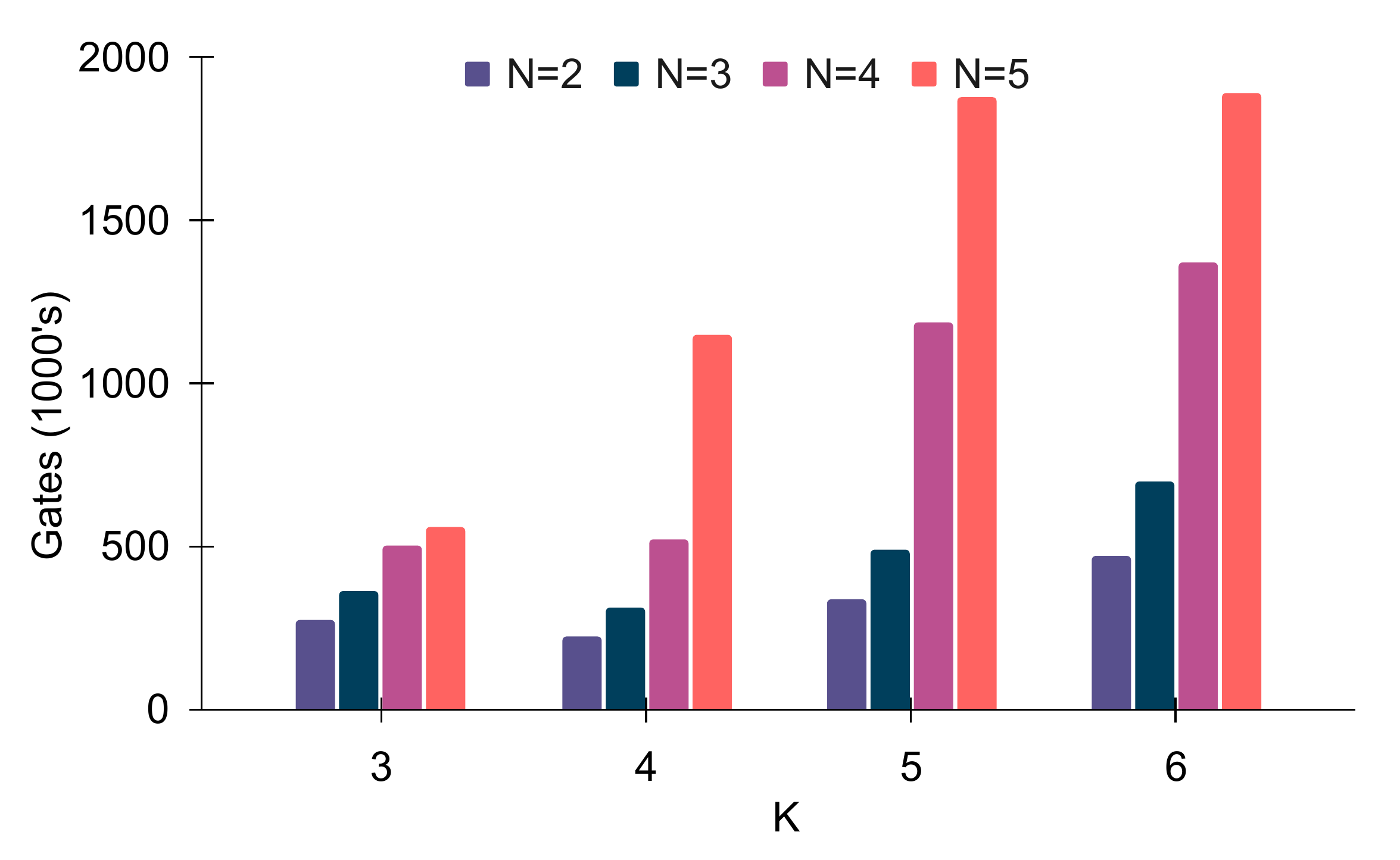}}
    }%
\caption{(a) Attack time and \# of gates in the unrolled 4$\times$4 fabric for N = 2. The attack timed-out for K = 7. (b) \# of gates in the unrolled 4$\times$4 fabric for various K, N.\label{fig:var_k}}
\end{figure}

\noindent \autoref{tab:full_icysat_4x4} gives the attack-time for different  $K$ and $N$ in a 4$\times$4 tile configuration, and the corresponding bitstream sizes, fabric size (measured as number of gates after producing an equivalent fabric netlist using 2-input gates for the attack), and the number of variables/clauses in the Boolean formula for the attack as reported by the \texttt{KC2} attack tool on a successful attack.  We set a time-out of 2 days.  All 6$\times$6 and all but one 5$\times$5 fabrics timed-out (smallest 5$\times$5 fabric, K3N2, bitstream size: 1204, was recovered in 18190 s). 

Focusing on the data from the successful attacks, the unroll factor does not show a monotonous or significant change as one varies $K$ while keeping $N$ constant.  This suggests that there is no significant progress / change in the complexity of combinational cycles in increasing the size of \acp{LUT} in CLB.  A higher $K$  adds higher complexity to local routing while maintaining or decreasing complexity of global routing. 
In contrast, increasing $N$ value while maintaining constant $K$ was found to cause a significant monotonic increase in unroll factor for distinct constant $K$ values. Increasing $N$ with constant $K$ significantly increased the global routing complexity, while maintaining the local routing complexity.  From these experiments, we can infer that the unroll factor, and hence the SAT attack complexity, is related to the complexity of the global routing network. 

\autoref{fig:fixed_k} shows the attack times for different values of $N$, with a fixed $K$=3 (note that $N$=6 and $N$=7 timed-out). These are plotted against the size of the unrolled fabric. 
This validates our claim that attack time increases with gate-size of the unrolled eFPGA fabric netlist.  Although the attack timed out for $K>3$, we expect the attack complexity trend for $K>3$ would continue, given the unrolled gate-size for $K>3$ as shown in \autoref{fig:GateN4x4}. 

\autoref{fig:fixed-N} shows the attack complexity as a function of $K$ with fixed $N$=2. 
As anticipated, the attack complexity increases with increasing $K$.  Increasing $K$ with fixed $N$ is associated with increasing SAT hardness due to presence of larger LUTs.  Although it also renders more complex local routing, previously we found that local routing does not appear to significantly change the unroll factor and hence does not contribute to the complexity arising from combinational cycles.  Hence we can say that the attack complexity in this case is primarily sourced from SAT hardness of LUTs.  For K=3, the attack time trend in \autoref{fig:fixed-N}  aligns with the trend seen for total 2 input gate-size shown in \autoref{fig:fixed-N}, which once again validates our claim that SAT attack complexity is directly related to the gate-size of unrolled eFPGA fabric. 

\autoref{fig:Gate-fixed-N-4x4} shows the total unrolled gate-size for various N, K values. 
The increase in SAT attack time-complexity due to increasing $N$ dominates vis-a-vis the increase in attack complexity due to increasing $K$.  This result hints at the fact that SAT attack complexity from cyclic networks within the eFPGA fabric contributes more to SAT attack complexity compared to the SAT attack complexity from SAT hardness of \acp{LUT} within the fabric.

\section{What Happens If We Partially Unroll?
\label{sec:security-II}}
In the previous section, we found that the complexity of the cyclic network and the associated SAT complexity of the eFPGA fabric is related to the complexity of global routing. 
The unroll factor required for a complete IcySAT attack is expected to increase significantly with more complex global routing. 
Since the unrolled gate-size of the eFPGA fabric is directly proportional to the size of the unroll factor, this is the primary quantifiable parameter to gauge SAT complexity.

\subsection{The Unroll Factor}
In Shamsi et al.'s original presentation of IcySAT~\cite{shamsi_icysat_2019}, the recommendation is to set the \texttt{unroll factor} equal to the cardinality of the non-optimal subset of nets that must be broken iteratively to remove the cycles in the circuit. 
With increasing complexity from cyclic networks in global routing of eFPGAs, more nets have to be broken to render an acyclic eFPGA netlist. 
For the general case, Shamsi {\em et al.} propose that a circuit should be unrolled \texttt{unroll factor} number of times to perfectly replicate the functions of a cyclic circuit with an acyclic equivalent. 
This is a based on a worst-case assumption that there might exist at least one trace between any pair of broken nets that traverses through all other broken nets, in which case, to perfectly replicate the intended acyclic behavior, the circuit has to be unrolled \texttt{unroll factor} times -- we will refer to this as the original IcySAT's ``ideal'' unroll factor. 
We contend that this worst-case scenario is a rare, at least in the context of eFPGA fabrics, which suggests that a partial unrolling might be sufficient for a successful attack. 
If we can unroll a circuit partially, the resulting Boolean formula will be smaller, and therefore more easily digested by a SAT solver. 

Thus, to investigate the possibility of recovering the bitstream after only unrolling a redaction fabric partially in the pre-processing step, we study the attack under three unroll factors: 10, 20 and 30. 
From the perspective of an adversary, the recovered bitstream is correct only if the locked netlist is found to be formally equivalent to the functional Oracle with the recovered bitstream applied. 
We follow the experimental setup in \autoref{sec:security-I}.

\subsection{Results}

\begin{table*}[ht]
\centering
\caption{Results from attacks using partially unrolled 4$\times$4 and 5$\times$5 eFPGA fabrics. The \% columns represent how much unrolling was performed relative to the ``ideal'' unroll factor used by the original IcySAT attack.}
\label{tab:partial_icysat}
\resizebox{\textwidth}{!}{%
\begin{tabular}{c c c c c c c c c c c c c c c }
\toprule
     & \multicolumn{7}{c}{4x4 Fabric}           & \multicolumn{7}{c}{5x5 Fabric}          \\ 
     \cmidrule(l){2-8} \cmidrule(l){9-15}
    & & & & \multicolumn{3}{c}{Unroll Factor} & & & & & \multicolumn{3}{c}{Unroll Factor} & \\ 
    \cmidrule(l){5-7} \cmidrule(l){12-14} 
\multirow{-3}{*}{Fabric} &   \multirow{-2}{*}{Time} &   \multirow{-2}{*}{Variables} &   \multirow{-2}{*}{Clauses} & 10 &   20 & 30 & \multirow{-2}{*}{\%} & \multirow{-2}{*}{Time} & \multirow{-2}{*}{Variables} & \multirow{-2}{*}{Clauses} & 10 & 20 & 30 &  \multirow{-2}{*}{\%} \\ \midrule
K3N2 & 14.2    & 259505  & 674076  & \xmark& \cmark & -- & 31 & 376.6   & 535654  & 1397911 & \xmark& \xmark& \cmark & 20\\ 
K3N3 & 27.6    & 317373  & 826296  & \xmark& \cmark & -- & 29 & 217     & 445372  & 1164011 & \xmark& \cmark& -- & 11\\ 
K3N4 & 53.1    & 260145  & 682636  & \xmark& \cmark & -- & 25 & 6066.7  & 822049  & 2170856 & \xmark& \xmark& \cmark & 19\\ 
K3N5 & 165.3   & 313473  & 827836  & \xmark& \cmark & -- & 27 & 79473.1 & 1148329 & 3053506 & \xmark& \xmark& \cmark & 13\\ 
K3N6 & 225.4   & 567108  & 1500081 & \xmark& \cmark & -- & 19 & 65301.6 & 1254310 & 3349951 & \xmark& \xmark& \cmark & 11\\ 
K3N7 & 48.13   & 222211  & 593346  & \cmark& -- & -- & 12   &  -- & --      & --      & \xmark& \xmark& \xmark & --\\ 
K3N8 & 1119.3  & 505917  & 1355976 & \xmark& \cmark & -- & 15   & --  & --      & --      & \xmark& \xmark& \xmark & --\\ 
K4N2 & 14.1    & 248639  & 628326  & \xmark& \cmark & -- & 36 & 162.8   & 421978  & 1099551 & \xmark& \cmark& -- & 13\\ 
K4N3 & FAIL    & --      & --      & \xmark& \xmark & \xmark & -- & 1298.1  & 544679  & 1433976 & \xmark& \cmark& -- & 14\\ 
K4N4 & 23.6    & 169281  & 446096  & \cmark& -- & -- & 16 & 9592.47 & 750851  & 1995976 & \xmark& \cmark& -- & 11\\ 
K4N5 & 265.6   & 414398  & 1097191 & \xmark& \cmark & -- & 18   & --  & --      & --      & \xmark& \xmark& \xmark & --\\ 
K4N6 & 926.1   & 503270  & 1344271 & \xmark& \cmark & -- & 18 & 25951.8 & 1153969 & 3093536 & \xmark& \cmark& -- & 7\\ 
K4N7 & 2915.2  & 593303  & 1601046 & \xmark& \cmark & -- & 24 & --      & --      & --      & \xmark& \cmark& -- & --\\ 
K4N8 & 5395    & 707405  & 1920016 & \xmark& \cmark & -- & 15 & --      & --      & --      & \xmark& \cmark& -- & --\\ 
K5N2 & 38.14   & 317687  & 820946  & \xmark& \cmark & -- & 31 & 378.2   & 534513  & 1397656 & \xmark& \cmark& -- & 12\\ 
K5N3 & 172.2   & 295576  & 774461  & \xmark& \cmark & -- & 29 & 2311    & 1086874 & 2868501 & \xmark& \xmark& \cmark & 14\\ 
K5N4 & 293.3   & 459865  & 1220356 & \xmark& \cmark & -- & 19 & --      & --      & --      & \xmark& \xmark& \xmark & --\\ 
K5N5 & 976     & 563633  & 1499516 & \xmark& \cmark & -- & 15 & --      & --      & --      & \xmark& \cmark& -- & 8\\ 
K5N6 & 105.5   & 351822  & 943831  & \cmark& -- & -- & 10  &   -- & --      & --      & \xmark& \xmark& \xmark &  -- \\ 
K5N7 & 10589.3 & 847415  & 2292606 & \xmark& \xmark & \cmark & 21  & --   & --      & --      & \xmark& \xmark& \xmark & --\\ 
K5N8 & 5582.6  & 999937  & 2722796 & \xmark& \cmark & -- &  12 & --   & --      & --      & \xmark& \xmark& \xmark & --\\ 
K6N2 & 71      & 279617  & 723756  & \xmark& \cmark & -- & 29 & 930.2   & 672571  & 1752806 & \xmark& \cmark& -- & 12\\ 
K6N3 & 32.85   & 205732  & 536881  & \cmark& -- & -- & 14 & 4130.3  & 978498  & 2583981 & \xmark& \cmark& -- & 11\\ 
K6N4 & 557.5   & 600741   & 1589936  &  \xmark&  \cmark&  -- & 22  &  -- & -- &  -- &  \xmark&  \xmark&  \xmark &  --\\ 
K6N5 & 3552    & 826572  & 2204601 & \xmark& \cmark & -- &  22  &   & --      & --      & \xmark& \xmark& \xmark & --\\ 
K6N6 & 4563.2  & 978410  & 2630371 & \xmark& \cmark & -- &   14 &   & --      & --      & \xmark& \xmark& \xmark & --\\ 
K6N7 & 170     & 1173923 & 3183426 & \cmark& -- & -- & 7   & --  & --      & --      & \xmark& \xmark& \xmark & --\\ 
K6N8 & 300.2   & 717045  & 1938196 & \cmark& -- & -- & 6   &   & --      & --      & \xmark& \xmark& \xmark & --\\ 
K7N2 & 68.6    & 389775  & 1006366 & \xmark& \cmark & -- & 37 & 1697.1  & 945518  & 2460251 & \xmark& \cmark& -- & 10\\ 
K7N3 & 45      & 303460  & 786641  & \cmark& -- & -- & 11 & 4376    & 1449330 & 3792581 & \xmark& \cmark& -- & 9\\ 
K7N4 & 1426.3  & 864697  & 2278516 & \xmark& \cmark & -- & 24      & -- & --      & --      & \xmark& \xmark& \xmark & --\\ 
K7N5 & 172.4   & 555561  & 1437396 & \cmark& -- & -- & 10      & --      & -- & --      & \xmark& \xmark& \xmark & \\ 
K7N6 & 3356    & 1380126 & 3693231 & \xmark& \cmark & -- & 14      & -- & --      & --      & \xmark& \xmark& \xmark & --\\ 
K7N7 & 443.9   & 1699415 & 1938196 & \cmark& -- & -- & 7      & -- & --      & --      & \xmark& \xmark& \xmark & --\\ 
K7N8 & 28407.7 & 2089178 & 5644891 & \xmark& \cmark & -- & 11      & -- & --      & --      & \xmark& \xmark& \xmark & --\\ \bottomrule
\end{tabular}%
}
\end{table*}

\noindent \autoref{tab:partial_icysat} presents the attack results on partially unrolled variants of 4$\times$4 and 5$\times$5 eFPGA fabrics. 
The attack was incrementally performed for different unroll factors until we recovered a bitstream that rendered the locked fabric netlist formally equivalent to the Oracle netlist. 
The (\xmark) in the table represents a failed attempt in which the recovered bitstream rendered a nonequivalent fabric, whereas the (\cmark) represents a correct bitstream solution. 
Upon getting a correct solution, further unrolling is skipped which is represented by (--). 
We observed that the bitstream for most of the fabrics in 4$\times$4 variants could be successfully recovered by partial unrolling, in contrast to the results of \autoref{sec:security-I}, 

The \% column in \autoref{tab:partial_icysat} shows that the adversary could successfully recover the bitstream, even when the unroll factor is as low as 6\% compared to the ``ideal'' unroll factor that would be used in the original IcySAT attack formulation. 
This demonstrates that the ``actual'' SAT complexity imparted by the cyclic networks of eFPGAs is lower than the expected complexity suggested by using the typical IcySAT algorithm. 
Although we do not have data representing the minimum unroll factor required to correctly recover the bitstream for each fabric, the data collected from successful attacks in this partial case might suggest that fabrics with higher ``ideal'' unroll factor (as determined by the original IcySAT algorithm) but smaller circuit size (in terms of the number of 2-input gates in the netlist) are more attack resilient compared to fabrics with lower ``ideal'' unroll factors but larger circuit size. 
In any case, the fact the bitstreams for so many of our fabrics were recovered in the partial unrolling case does raise concerns about the security of eFPGA-based redaction and merits further study. 


\section{Discussion\label{sec:discussion}\label{sec:practical}}



\begin{figure}[t!]
\centering
\subfloat[\label{fig:time_area}Attack time vs Area]{\includegraphics[width=0.5\columnwidth]{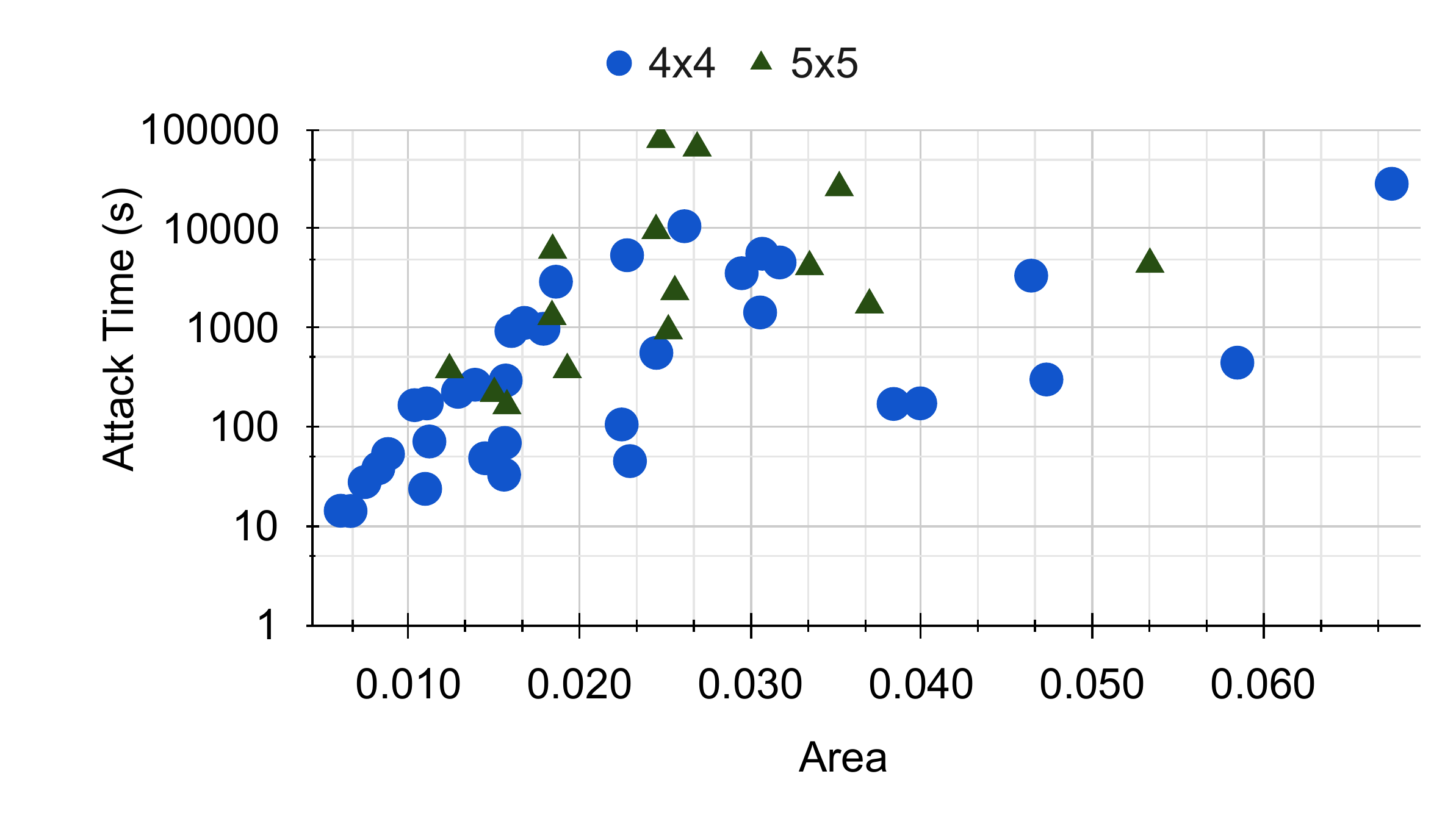}} 
\subfloat[\label{fig:time_area_delay}Attack time vs Area-Delay]{\includegraphics[width=0.5\columnwidth]{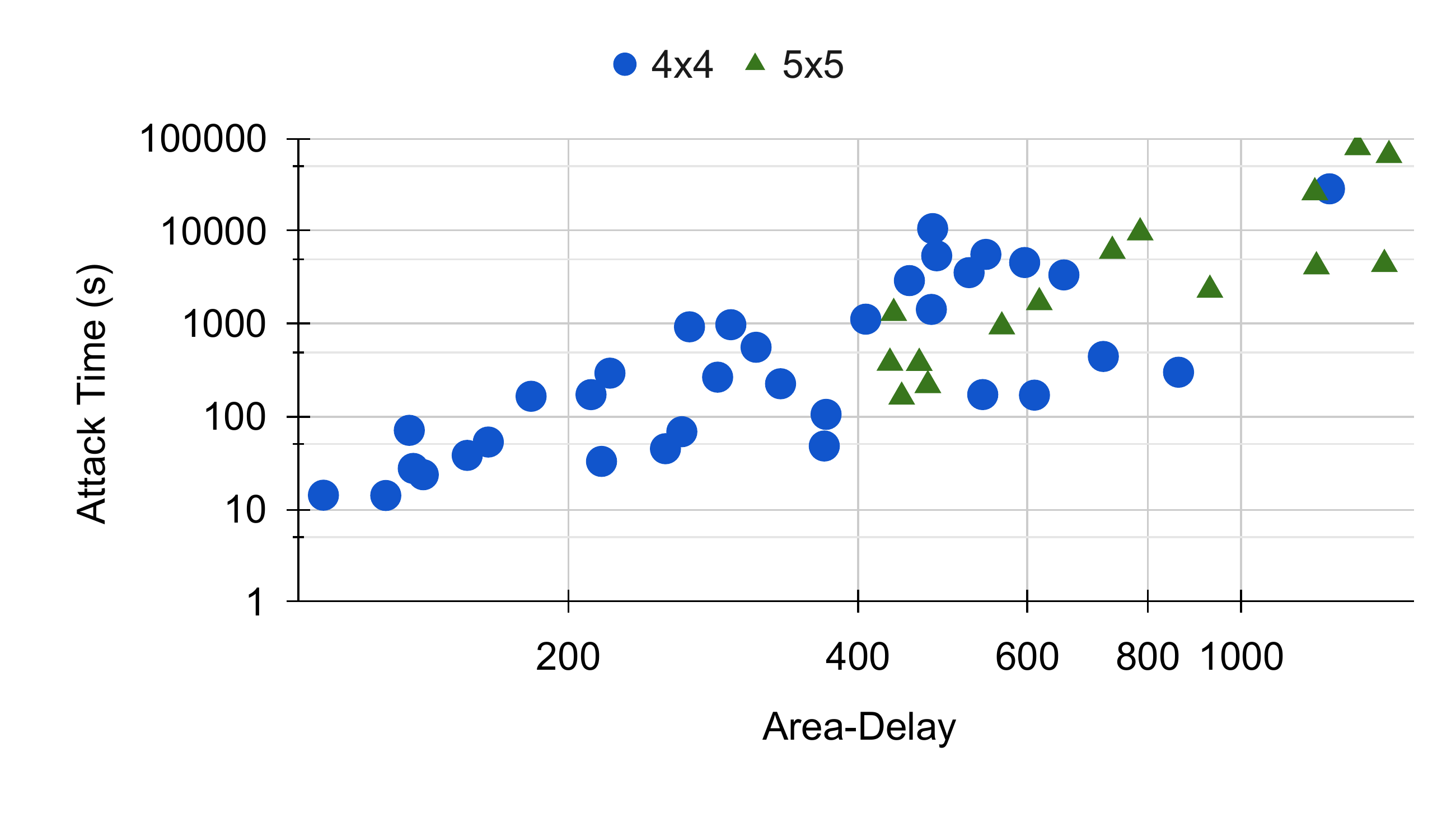}} 
\\

\subfloat[\label{fig:bits_total}Attack-time vs Bitstream]{\includegraphics[width=0.5\columnwidth]{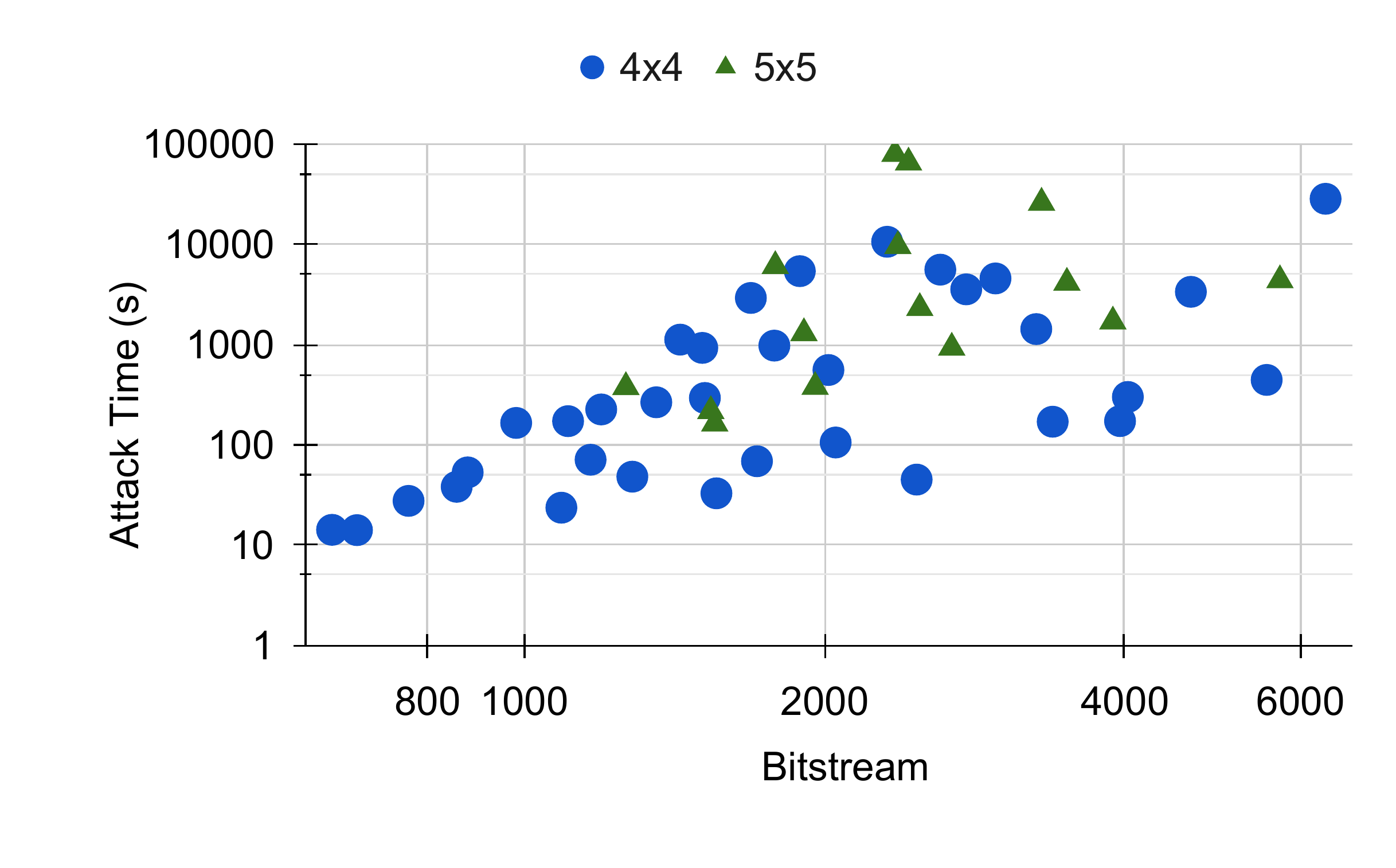}}
\subfloat[\label{fig:bits_logic}Attack-time vs Logic(Bitsream)]{\includegraphics[width=0.5\columnwidth]{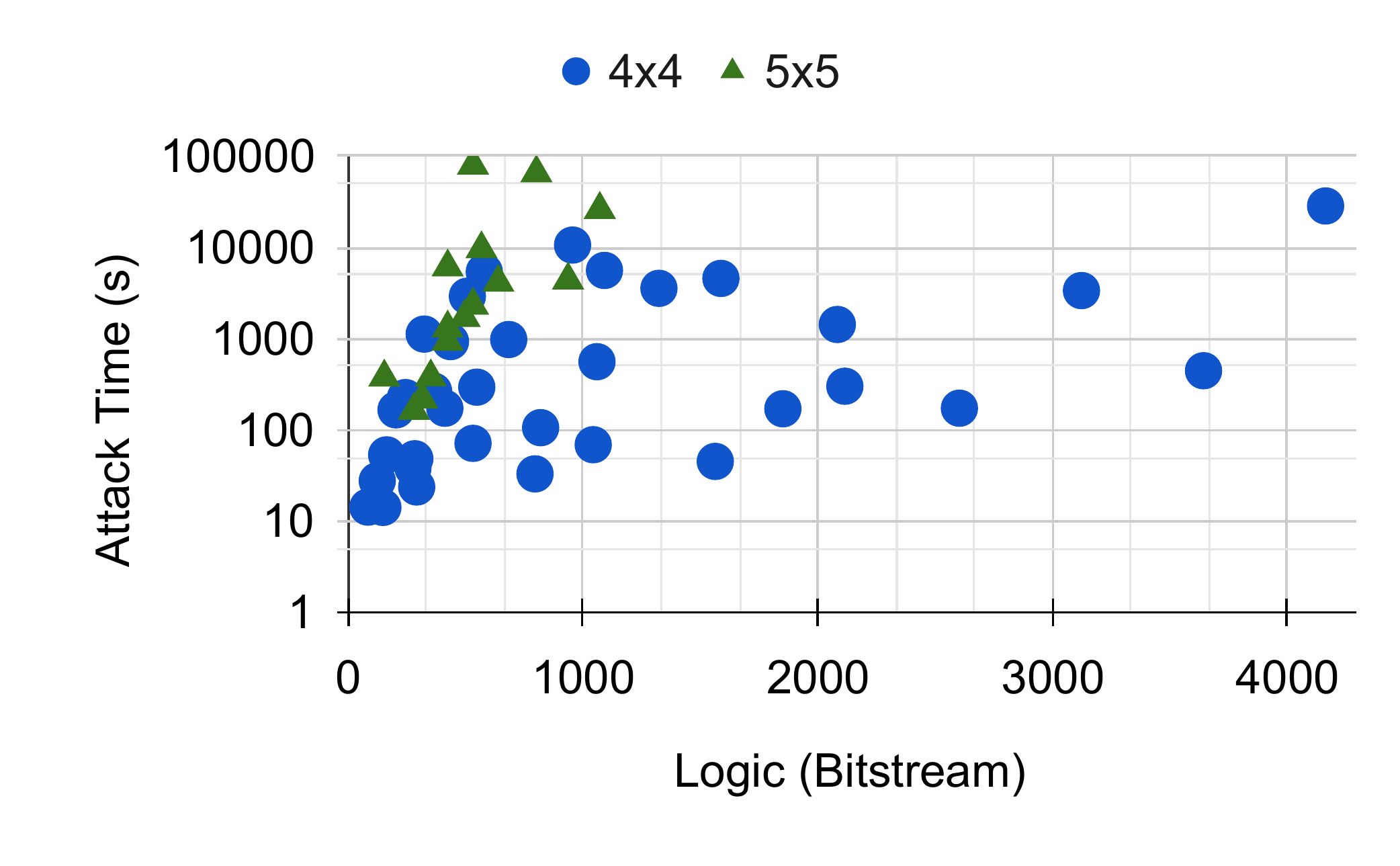}} 
\\
\subfloat[\label{fig:bits_local}Attack time vs LR (Bitstream)]{\includegraphics[width=0.5\columnwidth]{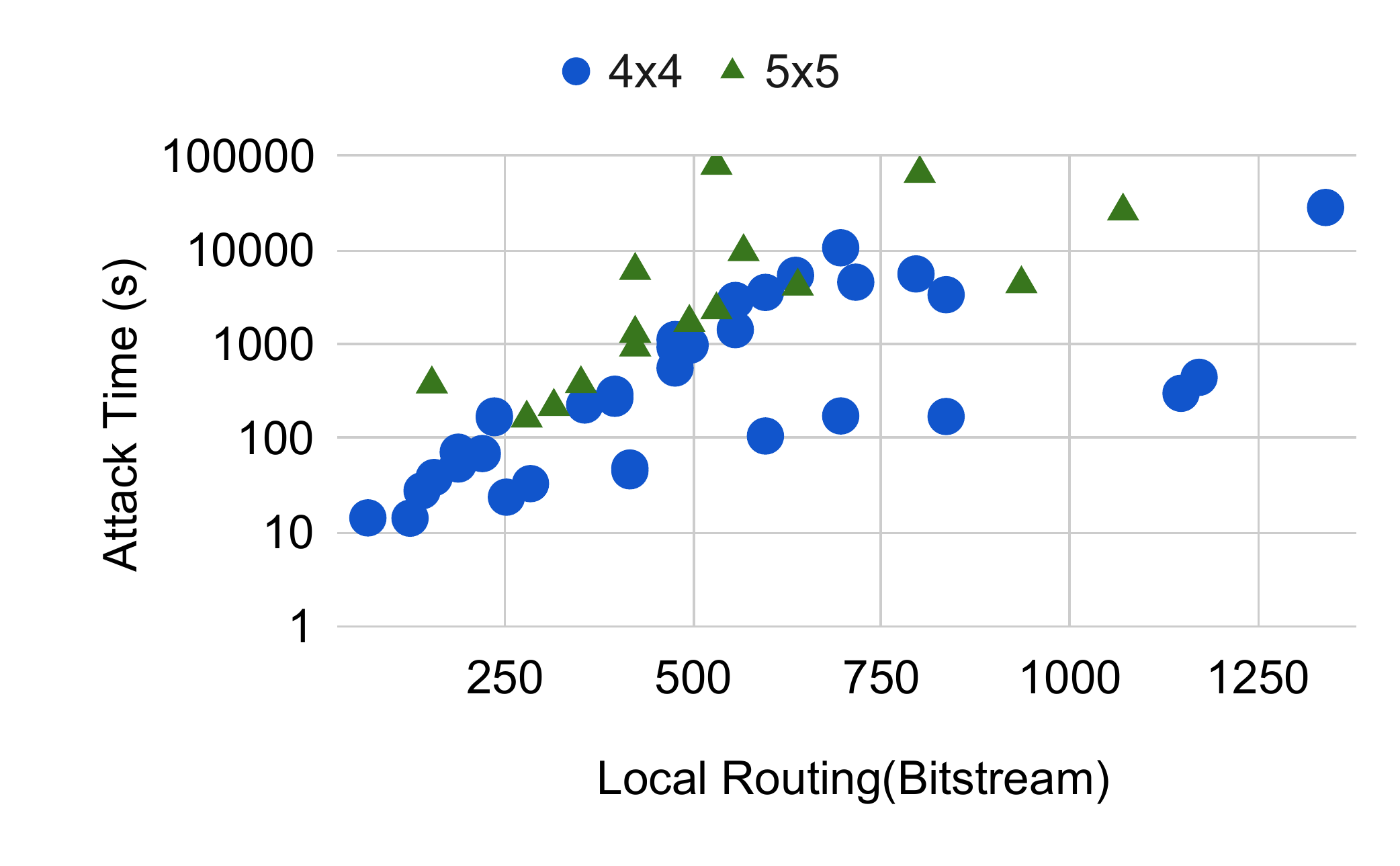}} \hfill
\subfloat[\label{fig:bits_global}Attack time vs GR (Bitstream)]{\includegraphics[width=0.5\columnwidth]{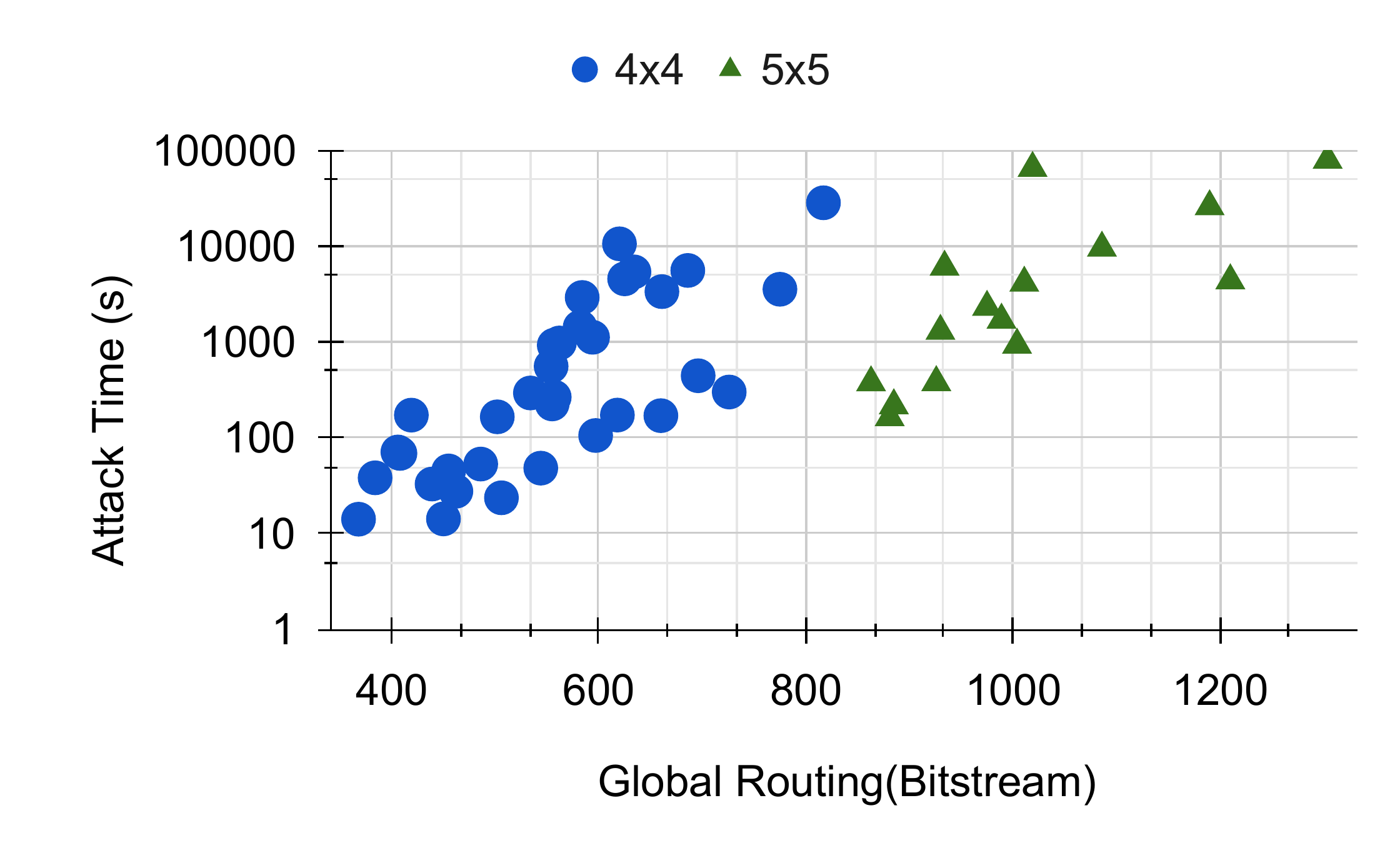}}
\caption{Characteristics of different parameters of eFPGA, vis-a-vis SAT-attack time. }
\label{fig:plot}
\end{figure}

\textbf{Gauging Security:} Given the results of our security assessment in \autoref{sec:security-I}, we now revisit the question: can we gauge security by considering only the bitstream size? 
As we alluded earlier, the bitstream, being the ``key'' in eFPGA redaction, might be thought of as the security parameter. 
This is the case in prior work \cite{mohan_hardware_2021} and other attempts at logic locking/obfuscation. 
In the case of eFPGA-based redaction, however, our experimental results indicate a more complex picture.  
Fig \ref{fig:bits_total} depicts the attack times for fabrics of different bitstream size. 
Although we can observe that the attack time generally increases with larger bitstream size, the distribution is scattered and hence the relationship between security and bitstream size is not definitive. 
One can argue that increasing the number of LUTs and LUT inputs makes SAT attacks harder; however, given that designers must also consider PPA overheads, our findings show that the story is more complex when also considering security.
We note that the attack time appears better correlated with the total number of gates in the unrolled netlist, which is similar to the trend observed in the product of unroll factor and the gate size of the netlist (\autoref{sec:security-I}). 
For instance, K5N3 with a bitstream size of 1066 is found to have approximately 10$\times$ attack time compared to K6N2 with a bitstream size of 1125. 
When we examine gate size however, notice that the gate size of the K5N3 fabric is significantly more than that of K6N2, resulting in more clauses for the SAT attack. 
This potentially explains the difference in attack time. 

To further explore the possible contribution of bitstream towards attack complexity, we examined the bitstream in terms of the number of bits used to configure the different parts of the of eFPGA. 
Since the eFPGAs used for redaction have fixed IO configuration bits, the overall bitstream could be categorized into three parts: (1) Logic configuration bits that set the contents of LUTs within the CLB; (2) Local routing configuration bits that select the input of the crossbar multiplexer that multiplex the \ac{CB} outputs to the LUTs; and (3) Global routing configuration bits, being the sum of \ac{CB} configuration bits and the \ac{SB} configuration bits. 
\autoref{fig:bits_logic}, \autoref{fig:bits_local}, and \autoref{fig:bits_global} (where, \textit{LR} and \textit{GR} stands for Local Routing and Global Routing respectively) demonstrate how the attack time varies with logic, local routing, and global routing configuration bits. 


\textbf{eFPGA Fabric Resource Utilization:} The cost associated with the redaction techniques tremendously increases as one move towards a more complex or different fabric size as shown in \autoref{tab:area_result} and \autoref{tab:power_result}. This ensures that a designer should have an idea of how much resources in terms of logic and I/Os are available in a fabric. This will lead to a better resource utilization in the fabric when one redacts a module, especially if one adopts a High Level Synthesis (HLS)-based ``top-down'' approach ~\cite{chen_decoy_2020, hu_functional_2019}. There can be two cases which limits the choice of a fabric: (1) Logic: This constitute the number of CLBs required to map a design; (2) I/Os: The number of inputs and outputs of a modules. For the first point, one can try to increase either \textit{K} and \textit{N} values in a fixed fabric size, to increase its logic capacity rather than moving to some bigger fabric sizes. To increase the number of I/Os, one can simply increase the capacity of the I/O tiles, but this comes at a cost greater routing complexity, but lower increase in overhead compared to the next sized fabric. 

\textbf{Area vs Security:} Our results from \autoref{tab:full_icysat_4x4} and \autoref{tab:partial_icysat} suggest that the security of a fabric is dependent on multiple parameters like fabric size, unroll factor, and their relative measures. We have shown that, even within a fixed fabric size, by varying \textit{K} and \textit{N}, the resulting SAT-based attack duration can vary considerably. Hence from a designer's perspective with a fixed area budget, a smart choice is to chose a fabric with the right size of resources while maximizing security by considering the insights from \autoref{sec:security-I} and \autoref{sec:security-II}. For better visualization, we have plotted the time taken by the successful attacks in against the area of the attacked fabric, shown in \autoref{fig:time_area}, where one can observe that, for a particular fixed value of area, the attack times can be significantly different for different fabric configuration. This implies that the security of a fabric is not solely dependant on the area, but on the combination of the various parameters of the FPGA architecture.
    
\textbf{Area-delay vs Security:} The Area-delay product provides a more general measure of the trade-off between area and delay~\cite{LUT}. 
As shown in \autoref{tab:delay_result}, delay decreases with increasing $K$ and $N$ as more logic can be clustered in one CLB and there are fewer paths that need to traverse through inter-cluster routing (CBs and SBs), where the delay will be quite large compared to intra-cluster routing. We have seen similar sort of results when we consider attack-time vs area-delay in \autoref{fig:time_area_delay}, where one can select a fabric with better area-delay and security parameters. 

\textbf{Study Limitations and Future Work:} 
Our study explores the security implications of eFPGA fabrics by varying two parameters: $K$ and $N$ from \autoref{tab:arch_param}.  The study in this paper motivates further scrutiny of eFPGA architectures for redaction to better enable better trade-offs between PPA and security. There are more parameters that can be configured in redaction fabric design as our future work. This includes alternative LUT designs, such as 
\textit{fracturable LUTs}, which have been shown to facilitate better resource utilization as more than one function can be mapped to a LUT with the cost of only a few  gates to separate the different outputs of a LUT.  This will somewhat increase the routing complexity (both local and global), with more numbers of input and outputs to consider, but could result in only a slight change in the overhead compared to mapping two different functions to separate LUTs in a conventional FPGA design~\cite{LUT,FPGA_arch}. 
In future work, we will extend the analysis to include BRAMs and DSPs as these elements in the fabric represent more points in the design space~\cite{FPGA_arch}.

For our study we fixed $W$, {$F_{c,in}$}, {$F_{c,out}$}, {$F_s$} and {$L$}. 
These present knobs that further expand the design space, and their implications on security should also be studied. 
%
With regards to our security evaluation, our partially-unrolled IcySAT experiments featured a limited number of unroll factors. 
In future work, we will find the minimum unroll factor at which a given bitstream can be recovered by more comprehensively sweeping unroll factors. 
Moreover, for designs where all 3 unroll factors failed to recover the correct bitstream, note that higher unroll factors and higher time-outs may still recover the correct keys. 
For instance, the partially unrolled IcySAT could successfully recover the keys for K4N3 variant of 4x4 eFPGA fabric for an unroll factor of 40. 
Hence wider sweep of unroll factor is required to evaluate the security of fabrics more thoroughly.


\section{Conclusions\label{sec:conclusions}}
This study has presented some of the key characteristics and security inferences of eFPGA architecture that have to be considered while performing redaction-based logic obfuscation. The study was performed by analyzing architectural variants of an eFPGA fabric by varying two parameters of an eFPGA: \textit{K} and \textit{N} which signifies the size and number of LUTs in the fabric. This gave several inferences on how security parameters relate to the architectural parameters. We framed a SAT-based security framework to recover the bitstream of FPGA that used the state-of-art IcySAT attack algorithm \cite{shamsi_icysat_2019} in the backend.  We experimentally concluded that security offered by an eFPGA fabric is primarily sourced from the SAT hardness of LUTs and the cyclic routing networks within the fabric. In contradiction to the assumptions from earlier work which stated that security was directly related to fabric size, we proved that in addition to fabric size, the attack complexity depends upon \textit{unroll factor}, a parameter of the IcySAT algorithm. We further showed that the primary contribution to an increasing unrolling factor is sourced from the complexity of global routing. We improvised the attack models to verify that existing attack parameters like unroll factor might not reflect the actual security strength of the process. Experiments showed that in most cases in 4x4 and 5x5 eFPGA fabrics, the adversary could recover the bitstream with an unroll factor of 7-36\% of the ideal unroll factor. Further, we experimentally disproved the assumption that bitstream size or size of any of its component is directly correlated to the security strength. We finally demonstrated how security strength might not strictly increase in proportion with physical parameters like area. 

Given that choice of fabric depends on individual circumstances, we cannot conclude that there is a single "good" or "best" fabric for redaction from our experiments. We show that bigger fabrics do not necessarily imply more SAT resilience; one can achieve a comparable level of security with a similar overhead. Thus, if a designer needs to consider overhead (which we think is typical), our results point to a need for co-designing to balance overheads and security.

\bibliographystyle{IEEEtran}
\bibliography{references}

\begin{thebibliography}{10}
\providecommand{\url}[1]{#1}
\csname url@samestyle\endcsname
\providecommand{\newblock}{\relax}
\providecommand{\bibinfo}[2]{#2}
\providecommand{\BIBentrySTDinterwordspacing}{\spaceskip=0pt\relax}
\providecommand{\BIBentryALTinterwordstretchfactor}{4}
\providecommand{\BIBentryALTinterwordspacing}{\spaceskip=\fontdimen2\font plus
\BIBentryALTinterwordstretchfactor\fontdimen3\font minus
  \fontdimen4\font\relax}
\providecommand{\BIBforeignlanguage}[2]{{%
\expandafter\ifx\csname l@#1\endcsname\relax
\typeout{** WARNING: IEEEtran.bst: No hyphenation pattern has been}%
\typeout{** loaded for the language `#1'. Using the pattern for}%
\typeout{** the default language instead.}%
\else
\language=\csname l@#1\endcsname
\fi
#2}}
\providecommand{\BIBdecl}{\relax}
\BIBdecl

\bibitem{rostami_primer_2014}
M.~Rostami, F.~Koushanfar, and R.~Karri, ``A {Primer} on {Hardware} {Security}:
  {Models}, {Methods}, and {Metrics},'' \emph{Proceedings of the IEEE}, vol.
  102, no.~8, pp. 1283--1295, Aug. 2014.

\bibitem{pilato_assure_2021}
C.~Pilato, A.~B. Chowdhury, D.~Sciuto, S.~Garg, and R.~Karri, ``{ASSURE}: {RTL}
  {Locking} {Against} an {Untrusted} {Foundry},'' \emph{IEEE Transactions on
  Very Large Scale Integration (VLSI) Systems}, pp. 1--13, 2021.

\bibitem{LLC}
B.~Tan \emph{et~al.}, ``Benchmarking at the frontier of hardware security:
  Lessons from logic locking,'' \emph{CoRR}, vol. abs/2006.06806, 2020.

\bibitem{shamsi_ip_2019}
K.~Shamsi, M.~Li, K.~Plaks, S.~Fazzari, D.~Z. Pan, and Y.~Jin, ``{IP}
  {Protection} and {Supply} {Chain} {Security} through {Logic} {Obfuscation}:
  {A} {Systematic} {Overview},'' \emph{ACM Transactions on Design Automation of
  Electronic Systems}, vol.~24, no.~6, pp. 65:1--65:36, Sep. 2019.

\bibitem{mohan_hardware_2021}
P.~Mohan, O.~Atli, J.~Sweeney, O.~Kibar, L.~Pileggi, and K.~Mai,
  ``\BIBforeignlanguage{en}{Hardware {Redaction} via {Designer}-{Directed}
  {Fine}-{Grained} {eFPGA} {Insertion}},'' in
  \emph{\BIBforeignlanguage{en}{Design, {Automation} \& {Test} in {Europe}
  {Conference} \& {Exhibition} ({DATE})}}.\hskip 1em plus 0.5em minus
  0.4em\relax Virtual: IEEE, 2021, p.~6.

\bibitem{chen_decoy_2020}
J.~Chen, M.~Zaman, Y.~Makris, R.~D.~S. Blanton, S.~Mitra, and B.~C. Schafer,
  ``{DECOY}: {DEflection}-{Driven} {HLS}-{Based} {Computation} {Partitioning}
  for {Obfuscating} {Intellectual} {PropertY},'' in \emph{2020 57th
  {ACM}/{IEEE} {Design} {Automation} {Conference} ({DAC})}.\hskip 1em plus
  0.5em minus 0.4em\relax San Francisco, CA, USA: IEEE, Jul. 2020, pp. 1--6.

\bibitem{hu_functional_2019}
B.~Hu \emph{et~al.}, ``\BIBforeignlanguage{en}{Functional {Obfuscation} of
  {Hardware} {Accelerators} through {Selective} {Partial} {Design} {Extraction}
  onto an {Embedded} {FPGA}},'' in \emph{\BIBforeignlanguage{en}{Proceedings of
  the 2019 on {Great} {Lakes} {Symposium} on {VLSI}}}.\hskip 1em plus 0.5em
  minus 0.4em\relax Tysons Corner VA USA: ACM, May 2019, pp. 171--176.

\bibitem{kamali_interlock_2020}
H.~M. Kamali, K.~Z. Azar, H.~Homayoun, and A.~Sasan,
  ``\BIBforeignlanguage{en}{{InterLock}: an intercorrelated logic and routing
  locking},'' in \emph{\BIBforeignlanguage{en}{Proceedings of the 39th
  {International} {Conference} on {Computer}-{Aided} {Design}}}.\hskip 1em plus
  0.5em minus 0.4em\relax Virtual Event USA: ACM, Nov. 2020, pp. 1--9.

\bibitem{kolhe_security_2019}
G.~Kolhe \emph{et~al.}, ``Security and {Complexity} {Analysis} of {LUT}-based
  {Obfuscation}: {From} {Blueprint} to {Reality},'' in \emph{2019 {IEEE}/{ACM}
  {International} {Conference} on {Computer}-{Aided} {Design} ({ICCAD})}.\hskip
  1em plus 0.5em minus 0.4em\relax Westminster, CO, USA: IEEE, Nov. 2019, pp.
  1--8.

\bibitem{liu_embedded_2014}
B.~Liu and B.~Wang, ``\BIBforeignlanguage{en}{Embedded reconfigurable logic for
  {ASIC} design obfuscation against supply chain attacks},'' in
  \emph{\BIBforeignlanguage{en}{Design, {Automation} \& {Test} in {Europe}
  {Conference} \& {Exhibition} ({DATE}), 2014}}.\hskip 1em plus 0.5em minus
  0.4em\relax Dresden, Germany: IEEE, 2014, pp. 1--6.

\bibitem{limaye_thwarting_2020}
N.~Limaye, E.~Kalligeros, N.~Karousos, I.~G. Karybali, and O.~Sinanoglu,
  ``Thwarting {All} {Logic} {Locking} {Attacks}: {Dishonest} {Oracle} with
  {Truly} {Random} {Logic} {Locking},'' \emph{IEEE Transactions on
  Computer-Aided Design of Integrated Circuits and Systems}, pp. 1--1, 2020.

\bibitem{subramanyan_evaluating_2015}
P.~Subramanyan, S.~Ray, and S.~Malik, ``Evaluating the security of logic
  encryption algorithms,'' in \emph{2015 {IEEE} {International} {Symposium} on
  {Hardware} {Oriented} {Security} and {Trust} ({HOST})}, May 2015, pp.
  137--143, iSSN: null.

\bibitem{shamsi_icysat_2019}
K.~Shamsi, D.~Z. Pan, and Y.~Jin, ``{IcySAT}: {Improved} {SAT}-based {Attacks}
  on {Cyclic} {Locked} {Circuits},'' in \emph{2019 {IEEE}/{ACM} {International}
  {Conference} on {Computer}-{Aided} {Design} ({ICCAD})}, Nov. 2019, pp. 1--7,
  iSSN: 1558-2434.

\bibitem{our_iccad}
J.~Bhandari \emph{et~al.}, ``Exploring \ac{eFPGA}-based redaction for \ac{IP}
  protection,'' in \emph{2021 {IEEE}/{ACM} {International} {Conference} on
  {Computer}-{Aided} {Design} ({ICCAD})}, Nov. 2021, p.~9.

\bibitem{FPGA_arch}
A.~Boutros and V.~Betz, ``Fpga architecture: Principles and progression,''
  \emph{IEEE Circuits and Systems Magazine}, vol.~21, no.~2, pp. 4--29, 2021.

\bibitem{tang_openfpga_2020}
X.~Tang, E.~Giacomin, B.~Chauviere, A.~Alacchi, and P.-E. Gaillardon,
  ``{OpenFPGA}: {An} {Open}-{Source} {Framework} for {Agile} {Prototyping}
  {Customizable} {FPGAs},'' \emph{IEEE Micro}, vol.~40, no.~4, pp. 41--48, Jul.
  2020, conference Name: IEEE Micro.

\bibitem{Azar_Kamali_Homayoun_Sasan_2018}
K.~Z. Azar, H.~M. Kamali, H.~Homayoun, and A.~Sasan, ``Smt attack: Next
  generation attack on obfuscated circuits with capabilities and performance
  beyond the sat attacks,'' \emph{IACR Transactions on Cryptographic Hardware
  and Embedded Systems}, vol. 2019, no.~1, pp. 97--122, Nov. 2018.

\bibitem{be_sat}
Y.~Shen, Y.~Li, A.~Rezaei, S.~Kong, D.~Dlott, and H.~Zhou,
  ``\BIBforeignlanguage{en}{{BeSAT}: behavioral {SAT}-based attack on cyclic
  logic encryption},'' in \emph{\BIBforeignlanguage{en}{Proceedings of the 24th
  {Asia} and {South} {Pacific} {Design} {Automation} {Conference}}}.\hskip 1em
  plus 0.5em minus 0.4em\relax Tokyo Japan: ACM, Jan. 2019, pp. 657--662.

\bibitem{cyc_sat}
H.~Zhou, R.~Jiang, and S.~Kong, ``{CycSAT}: {SAT}-based attack on cyclic logic
  encryptions,'' in \emph{2017 {IEEE}/{ACM} {International} {Conference} on
  {Computer}-{Aided} {Design} ({ICCAD})}, Nov. 2017, pp. 49--56, iSSN:
  1558-2434.

\bibitem{shamsi_kc2_2019}
K.~Shamsi, M.~Li, D.~Z. Pan, and Y.~Jin, ``{KC2}: {Key}-{Condition} {Crunching}
  for {Fast} {Sequential} {Circuit} {Deobfuscation},'' in \emph{2019 {Design},
  {Automation} \& {Test} in {Europe} {Conference} \& {Exhibition}
  ({DATE})}.\hskip 1em plus 0.5em minus 0.4em\relax Florence, Italy: IEEE, Mar.
  2019, pp. 534--539.

\bibitem{li_piercing_2019}
L.~Li and A.~Orailoglu, ``Piercing {Logic} {Locking} {Keys} through
  {Redundancy} {Identification},'' in \emph{2019 {Design}, {Automation} {Test}
  in {Europe} {Conference} {Exhibition} ({DATE})}, Mar. 2019, pp. 540--545,
  iSSN: 1530-1591.

\bibitem{han_does_2021}
Z.~Han, M.~Yasin, and J.~J. Rajendran, ``Does logic locking work with {EDA}
  tools?'' in \emph{30th {USENIX} Security Symposium ({USENIX} Security
  21)}.\hskip 1em plus 0.5em minus 0.4em\relax {USENIX} Association, Aug. 2021,
  available:
  \url{https://www.usenix.org/conference/usenixsecurity21/presentation/han-zhaokun}.

\bibitem{Shakya_Xu_Tehranipoor_Forte_2019}
B.~Shakya, X.~Xu, M.~Tehranipoor, and D.~Forte, ``Cas-lock: A
  security-corruptibility trade-off resilient logic locking scheme,''
  \emph{IACR Transactions on Cryptographic Hardware and Embedded Systems}, vol.
  2020, no.~1, pp. 175--202, Nov. 2019.

\bibitem{Sengupta_Limaye_Sinanoglu_2021}
A.~Sengupta, N.~Limaye, and O.~Sinanoglu, ``Breaking cas-lock and its variants
  by exploiting structural traces,'' \emph{IACR Transactions on Cryptographic
  Hardware and Embedded Systems}, vol. 2021, no.~3, pp. 418--440, Jul. 2021.

\bibitem{tang_openfpga_2019}
X.~Tang, E.~Giacomin, A.~Alacchi, B.~Chauviere, and P.-E. Gaillardon,
  ``{OpenFPGA}: {An} {Opensource} {Framework} {Enabling} {Rapid} {Prototyping}
  of {Customizable} {FPGAs},'' in \emph{2019 29th {International} {Conference}
  on {Field} {Programmable} {Logic} and {Applications} ({FPL})}, Sep. 2019, pp.
  367--374, iSSN: 1946-1488.

\bibitem{flexlogic}
``{TSMC 40ULP \& 40LP EFLX 1K eFPGA Tile: GDS AVAILABLE},''
  \url{https://flex-logix.com/wp-content/uploads/2020/09/2020-09-EFLX-1K-TSMC-40ULP40LP-AVAILABLE-product-brief.pdf}.

\bibitem{LUT}
E.~Ahmed and J.~Rose, ``The effect of lut and cluster size on deep-submicron
  fpga performance and density,'' \emph{IEEE Transactions on Very Large Scale
  Integration (VLSI) Systems}, vol.~12, no.~3, pp. 288--298, 2004.

\bibitem{global_routing}
G.~Lemieux, E.~Lee, M.~Tom, and A.~Yu, ``Directional and single-driver wires in
  fpga interconnect,'' in \emph{Proceedings. 2004 IEEE International Conference
  on Field- Programmable Technology (IEEE Cat. No.04EX921)}, 2004, pp. 41--48.

\bibitem{koch_fabulous_2021}
D.~Koch, N.~Dao, B.~Healy, J.~Yu, and A.~Attwood,
  ``\BIBforeignlanguage{en}{{FABulous}: {An} {Embedded} {FPGA} {Framework}},''
  in \emph{\BIBforeignlanguage{en}{The 2021 {ACM}/{SIGDA} {International}
  {Symposium} on {Field}-{Programmable} {Gate} {Arrays}}}.\hskip 1em plus 0.5em
  minus 0.4em\relax Virtual Event USA: ACM, Feb. 2021, pp. 45--56.

\bibitem{mohan_top-down_2021}
P.~Mohan, O.~Atli, O.~Kibar, M.~Zackriya, L.~Pileggi, and K.~Mai,
  ``\BIBforeignlanguage{en}{Top-down {Physical} {Design} of {Soft} {Embedded}
  {FPGA} {Fabrics}},'' in \emph{\BIBforeignlanguage{en}{The 2021 {ACM}/{SIGDA}
  {International} {Symposium} on {Field}-{Programmable} {Gate}
  {Arrays}}}.\hskip 1em plus 0.5em minus 0.4em\relax Virtual Event USA: ACM,
  Feb. 2021, pp. 1--10.

\bibitem{intel_xeon_fpga}
``{Intel Xeon+FPGA Platform for the Data Center},''
  \url{https://reconfigurablecomputing4themasses.net/files/2.2\%20PK.pdf}.

\bibitem{PSchiavone_tvlsi_2021}
P.~D. Schiavone \emph{et~al.}, ``Arnold: {An} {eFPGA}-{Augmented} {RISC}-{V}
  {SoC} for {Flexible} and {Low}-{Power} {IoT} {End} {Nodes},'' \emph{IEEE
  Transactions on Very Large Scale Integration (VLSI) Systems}, vol.~29, no.~4,
  pp. 677--690, Apr. 2021.

\bibitem{ALi_FPGA_2021}
A.~Li and D.~Wentzlaff, ``\BIBforeignlanguage{en}{{PRGA}: {An} {Open}-{Source}
  {FPGA} {Research} and {Prototyping} {Framework}},'' in
  \emph{\BIBforeignlanguage{en}{The 2021 {ACM}/{SIGDA} {International}
  {Symposium} on {Field}-{Programmable} {Gate} {Arrays}}}.\hskip 1em plus 0.5em
  minus 0.4em\relax Virtual Event USA: ACM, Feb. 2021, pp. 127--137.

\bibitem{JLuu_FPGA_2011}
J.~Luu, J.~H. Anderson, and J.~S. Rose, ``{Architecture} {Description} and
  {Packing} for {Logic} {Blocks} with {Hierarchy}, {Modes} and {Complex}
  {Interconnect},'' in \emph{Proceedings of the 19th ACM/SIGDA International
  Symposium on Field Programmable Gate Arrays}, ser. FPGA '11.\hskip 1em plus
  0.5em minus 0.4em\relax New York, NY, USA: Association for Computing
  Machinery, 2011, p. 227–236.

\bibitem{XTang_tvlsi_2018}
X.~Tang, E.~Giacomin, G.~D. Micheli, and P.-E. Gaillardon, ``{FPGA}-{SPICE}:
  {A} {Simulation}-{Based} {Architecture} {Evaluation} {Framework} for
  {FPGAs},'' \emph{IEEE Transactions on Very Large Scale Integration (VLSI)
  Systems}, vol.~27, no.~3, pp. 637--650, Mar. 2019.

\bibitem{GGore_ispd_2021}
G.~Gore, X.~Tang, and P.-E. Gaillardon, ``\BIBforeignlanguage{en}{A {Scalable}
  and {Robust} {Hierarchical} {Floorplanning} to {Enable} 24-hour {Prototyping}
  for 100k-{LUT} {FPGAs}},'' in \emph{\BIBforeignlanguage{en}{Proceedings of
  the 2021 {International} {Symposium} on {Physical} {Design}}}.\hskip 1em plus
  0.5em minus 0.4em\relax Virtual Event USA: ACM, Mar. 2021, pp. 135--142.

\bibitem{CWolf_yosys_2013}
C.~Wolf, J.~Glaser, and J.~Kepler, ``Yosys-a free verilog synthesis suite,'' in
  \emph{Proceedings of Austrochip}, 2013.

\bibitem{KMurray_trets_2020}
K.~E. Murray \emph{et~al.}, ``\BIBforeignlanguage{en}{{VTR} 8:
  {High}-performance {CAD} and {Customizable} {FPGA} {Architecture}
  {Modelling}},'' \emph{\BIBforeignlanguage{en}{ACM Transactions on
  Reconfigurable Technology and Systems}}, vol.~13, no.~2, pp. 1--55, Jun.
  2020.

\bibitem{FreePDK45}
``{FreePDK45}:{Contents} - {NCSU} {EDA} {Wiki},''
  \url{https://www.eda.ncsu.edu/wiki/FreePDK45:Contents}.

\bibitem{trimberger_fpga_2014}
S.~M. Trimberger and J.~J. Moore, ``{FPGA} {Security}: {Motivations},
  {Features}, and {Applications},'' \emph{Proceedings of the IEEE}, vol. 102,
  no.~8, pp. 1248--1265, Aug. 2014, conference Name: Proceedings of the IEEE.

\bibitem{rahman_key_2020}
M.~T. Rahman, S.~Tajik, M.~S. Rahman, M.~Tehranipoor, and N.~Asadizanjani,
  ``The {Key} is {Left} under the {Mat}: {On} the {Inappropriate} {Security}
  {Assumption} of {Logic} {Locking} {Schemes},'' in \emph{2020 {IEEE}
  {International} {Symposium} on {Hardware} {Oriented} {Security} and {Trust}
  ({HOST})}.\hskip 1em plus 0.5em minus 0.4em\relax San Jose, CA, USA: IEEE,
  Dec. 2020, pp. 262--272.

\bibitem{LUT_LOCK}
H.~Mardani~Kamali, K.~Zamiri~Azar, K.~Gaj, H.~Homayoun, and A.~Sasan,
  ``Lut-lock: A novel lut-based logic obfuscation for fpga-bitstream and
  asic-hardware protection,'' in \emph{2018 IEEE Computer Society Annual
  Symposium on VLSI (ISVLSI)}, 2018, pp. 405--410.

\end{thebibliography}
\end{document}